\documentclass[sigplan,nonacm]{acmart}

\usepackage[]{hyperref}
    
\usepackage{balance}
\usepackage[utf8]{inputenc}
\usepackage{xspace}
\usepackage{subcaption}
\usepackage{listings}
\usepackage{algorithm}
\usepackage[noend]{algpseudocode}
\usepackage{xcolor}
\usepackage{array}
\usepackage{tabularx}
\usepackage{arydshln}
\usepackage{makecell}

\usepackage{rotating}

\usepackage{ifthen}
\usepackage{multirow, booktabs}
\usepackage{siunitx}
\usepackage{tikz}
\newcommand*\circled[1]{\tikz[baseline=(char.base)]{
            \node[shape=circle,draw,inner 
            sep=1pt,fill={rgb:yellow,2;white,1}] (char) {#1};}}

\newcommand{\CommentCL}[1]{\Comment{\color{gray}{#1}\color{black}}}
\algdef{SE}[SUBALG]{Indent}{EndIndent}{}{\algorithmicend\ }%
\algtext*{Indent}
\algtext*{EndIndent}

\newtheorem{property}{Property}

\newboolean{showcomments}
\setboolean{showcomments}{false}
\ifthenelse{\boolean{showcomments}}
{ \newcommand{\mynote}[3]{
    \fbox{\bfseries\sffamily\scriptsize#1}
    {\small\textsf{\emph{\color{#3}{#2}}}}}}
{ \newcommand{\mynote}[3]{}}

\newcommand{\jpb}[1]{\mynote{jpb}{#1}{orange}}

\title{DUMBO: Making durable read-only transactions \\fly on hardware transactional memory}

\author{João Barreto}
\affiliation{%
\institution{INESC-ID \& IST Universidade de Lisboa}
\country{Portugal}
}

\author{Daniel Castro}
\affiliation{%
\institution{INESC-ID \& IST Universidade de Lisboa}
\country{Portugal}
}

\author{Paolo Romano}
\affiliation{%
\institution{INESC-ID \& IST Universidade de Lisboa}
\country{Portugal}
}

\author{Alexandro Baldassin}
\affiliation{%
\institution{São Paulo State University (Unesp)}
\country{Brazil}
}

\makeatletter
\g@addto@macro{\UrlBreaks}{\UrlOrds}
\makeatother

\begin{document}

\newcommand{\system}{DUMBO\xspace}
\newcommand{\naive}{SPHT+SI-HTM\xspace}

\begin{abstract}

\jpb{The Compute Express Link (CXL) enables in-memory DBMSs to rely on new persistent memory devices. 
Hardware Transactional Memory (HTM) 
is a promising solution to support durable transactions on such emerging technology.
However, although recent advances in Persistent Hardware Transactions (PHTs)
have achieved important improvements on the scalability of update transactions,}
Despite the recent improvements in supporting Persistent Hardware Transactions (PHTs)
on emerging persistent memories (PM),
the poor performance of Read-Only (RO) transactions remains largely overlooked.
\jpb{they have largely overlooked the poor performance of Read-Only (RO) transactions, which
suffer from two crucial bottlenecks: considerable post-commit delays are required 
to ensure consistency with concurrent update transactions, and the well-known 
tight read capacity limits of the commercially available HTM implementations.}
We propose \system, a new design for PHT 
that eliminates the two most crucial bottlenecks that hinder RO transactions in state-of-the-art PHT. 
At its core, \system exploits advanced instructions that some contemporary HTMs
provide to suspend (and resume) transactional access tracking.
\jpb{Although \system adds a new synchronization step to update transactions, 
it also incorporates novel optimizations that reduce the durability overheads in 
update transactions. }
%
Our experimental evaluation with an IBM POWER9 system using the TPC-C benchmark 
shows that \system can outperform the state of the art designs for persistent 
hardware (SPHT) and software memory transactions (Pisces), by up to 4.0$\times$.

\end{abstract}

\maketitle

\pagestyle{plain}

\section{Introduction}
\label{sec:intro}

The emergence of the Compute Express Link (CXL) technology~\cite{OptaneCXL} has sparkled a profound shift towards 
high-capacity and high-throughput Persistent Memory (PM) devices.
With CXL 2.0, the first commercial PM devices are already under development%
~\cite{samsung:cxl} and, in a near future, disaggregated CXL 3.0-based PM~\cite{10.1145/3624062.3624175} will become a reality.
The rise of 
byte-addressable PM allows bridging the gap between 
volatile and scarce memory vs. persistent and large disk storage, which has guided decades of software \jpb{DBMS} design. 

\jpb{In recent years, many popular DBMSs have started to break free of such dichotomy, as their newer generations have begun to exploit the virtues of PM.
Notable examples include SAP HANA~\cite{SAPHANA_VLDB17}, SQL Server~\cite{SQLServerPMEM}, Oracle Exadata X8M~\cite{OracleExadataX8M}, MySQL~\cite{MySQLPMEM}, and PostgresSQL~\cite{PostgresqlPMEM}.
For a DBMS to operate correctly with PM, appropriate mechanisms 
must be put in place to address two fundamental requirements:
\emph{concurrency control} (to leverage the parallel power of multi-cores); and \emph{durability} 
(to ensure that, after a crash, the system is able to recover to a consistent state). }

At a first glance, the low synchronization overheads of Hardware Transactional Memory (HTM) 
\cite{SAPHANA_HPCA14} make this technology a promising solution for handling concurrent transactions 
accessing shared data that is durably stored in PM.
However,
existing HTM implementations do not  guarantee that updates generated by a committed  transaction are atomically 
transposed from the (volatile) CPU cache to PM
~\cite{ giles2017ISMM,CastroIPDPS2018,GencPLDI2020,Baldassin2020}.
A number of proposals of Persistent Hardware Transactions (PHT) extend off-the-shelf HTM implementations with software-based schemes, which ensure that the effects of hardware 
transactions on PM remain consistent in the presence of system crashes~\cite{LiuASPLOS2017,giles2017ISMM,CastroIPDPS2018,GencPLDI2020,SPHT}.
\jpb{Today's state of the art solutions in PHT maintain a write-ahead log in PM for each hardware 
transaction that is executing. Any updates that the transaction issues are performed on a shadow copy in DRAM. 
Once a transaction commits in HTM, the log is persisted to PM
and the transactions coordinates with concurrent transactions to ensure that the order by which
their effects become visible is consistent with the order by which they become durable in PM.
This ensures that each transaction's effects can be consistently recovered if the system crashes.}


\begin{figure}[t]
    \centering
    \includegraphics[width=0.40\textwidth]{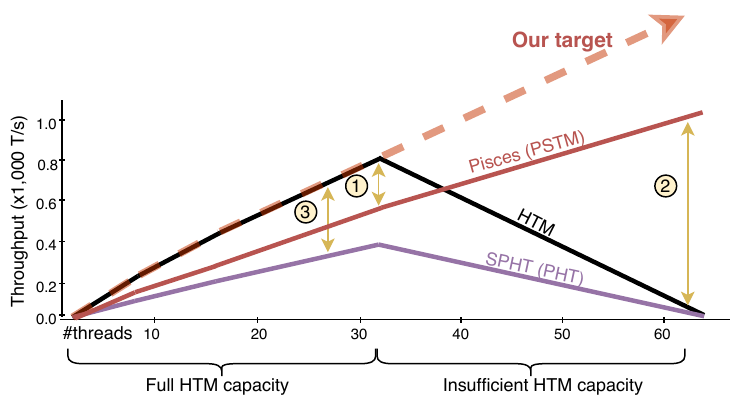}
    \vspace{-3mm}
    \caption{Throughput of RO transactions in a read-dominated TPC-C workload (\emph{Payment} transactions on 1 thread + 
    multiple threads running \emph{Order status} transactions).}
    \label{fig:motivation}
    \vspace{-3mm}
\end{figure}

Since read-only (RO) transactions are dominant in many real-world workloads, the design of 
PHT should prioritize their performance. 
However, while the state of the art in PHT 
has achieved important improvements on the scalability of update transactions~\cite{SPHT},
it has largely overlooked the poor performance of RO transactions.

To support this claim, we set out 
a simple experiment where a variable number of threads concurrently executes 
\emph{order status} RO transactions from the TPC-C benchmark~\cite{tpcc}, while a single thread  continuously
runs \emph{payment} update transactions. 
We use this workload to study three representative systems: 
SPHT, the state of the art on PHT~\cite{SPHT}; 
Pisces, a read-optimized Persistent Software TM (PSTM)~\cite{Gu2019ATC}; 
and  plain HTM, as a non-durable baseline.
We use an IBM POWER9 machine with a total of 32 cores, supporting HTM and Simultaneous Multithreading (SMT)
(for detailed specifications, 
see \S\ref{sec:evaluation}).
Up to 32 threads, each thread runs alone on a dedicated core; 
beyond that, thread pairs are co-located in the same core, which roughly halves the 
per-thread transactional capacity.

Figure~\ref{fig:motivation} shows the RO transaction throughput of each solution, as
we increase the number of RO threads.
Despite the promises of HTM-based solutions, their performance is disappointing, while PSTM 
 is the most competitive.
This can be explained by three main observations, depicted in the plot.
 \circled{1} HTM is faster than STM when running with enough capacity (up to 32 threads). 
\circled{2} As soon as capacity drops, read sets no longer fit in HTM; thus, HTM-based solutions start thrashing 
with capacity aborts and falling back to the
Single Global Lock (SGL) path, a well-known limitation of HTM \cite{castro2017MASCOTS}.
\jpb{This is a well-known limitation HTM, 
which is at odds with the large footprints of typical real-world OLTP/OLAP workloads~\cite{castro2017MASCOTS,FilipePPoPP2019,Baldassin2020}.}
\circled{3}
Even in the HTM-friendly zone (up to 32 threads), 
SPHT imposes a substantial overhead, 
which makes it less competitive than STM. 
In fact, 
to guarantee consistency in the presence of crashes \cite{durableopacity20}, any RO transaction, $R$, that completes its execution
must wait until any write that $R$ might have observed is already durable. 
In SPHT, this is enforced by a \emph{durability wait} that only terminates after 
any transaction that either executed before or concurrently with $R$ has become durable.


%
The main goal of this paper is to eliminate the fundamental bottlenecks of RO transactions in 
PHT, thereby achieving ``our target'' in Figure \ref{fig:motivation}.
We take advantage of instructions for suspending (and resuming) access 
tracking inside hardware transactions, which are offered by commercial HTM implementations. 
We draw inspiration from existing proposals that exploit such instructions to 
relieve hardware transactions from the read capacity bounds of 
HTM~\cite{FilipePPoPP2019}.
Their approach is to suspend load tracking, which grants the transaction unlimited read capacity.
To prevent a potential breach of isolation, 
update transactions 
perform an \emph{isolation wait} before they can safely commit in HTM.

\textbf{As a first contribution}, 
we empirically demonstrate that, although we can directly combine existing designs for PHT with proposed techniques 
for unlimited reads in hardware transactions,
a naive combination of both approaches
incurs prohibitive overheads. 

These findings motivate our \textbf{second contribution}, the design of \emph{Du\-ra\-ble Unlimited-reads Memory transactions on Best-effOrt HTM (\system)}.
\system relies on two main building blocks.
A first one is a novel \emph{RO durability wait} that
\jpb{based on the premise that update transactions now run the \emph{isolation wait} before 
committing in HTM. 
In this revised wait logic,} spares RO transactions from the need to wait for concurrent update transactions 
to become durable.
In practice, this optimization reduces the RO durability overheads to a negligible level.
Together, the unlimited reads feature and our proposed \emph{RO durability wait} allow 
\system to overcome the two key limitations depicted in Figure \ref{fig:motivation}, 
hence fulfilling the main goal of this paper.

Both advantages are only possible because our solution sacrifices update transactions with a 
time-consuming \emph{isolation wait} phase before HTM commit. 
Hence, a second building block comprises a set of new techniques (\emph{opportunistic redo log flushing}  and \emph{partially-ordered durability markers}) that exploit the 
\emph{isolation wait} as an opportunity to rethink the expensive durability stages of update transactions.
\jpb{Concretely: \emph{opportunistic redo log flushing} hides the latency of log flushing by 
overlapping it with the \emph{isolation wait}; 
and \emph{partially-ordered durability markers}, which, by leveraging a property ensured by the isolation wait, dramatically reduces the post-commit coordination that transactions must carry on before becoming durable.}
With these optimizations, \jpb{, once an update transaction leaves the \emph{isolation wait} and commits in HTM, }
the durability steps of update transactions in \system become substantially faster than in SPHT.
This compensates for the additional latency of in \system's \emph{isolation wait} and, in
some workloads, even outweighs it.

\system can ensure either 
opacity~\cite{opacity} (the established correctness criteria in transactional memory literature) or
the weaker Snapshot Isolation (SI)~\cite{Berenson1995SIGMOD,Cerone2016PODC,Fekete2015}.
This choice entails a trade-off: with opacity, only RO transactions benefit from unlimited reads; 
with SI, all transactions do.


\textbf{As a third contribution}, we implement and experimentally evaluate \system in an IBM POWER9 system.
\jpb{, 
emulating the access latency of the upcoming CXL-based PM devices.}
Using diverse workloads from the TPC-C benchmark \cite{tpcc}, we compare 
\system with two systems that, to our knowledge, are among the most
competitive proposals for durable memory transactions on PM:  
 SPHT \cite{SPHT} (PHT) and Pisces \cite{Gu2019ATC} (PSTM).
Our results show that, at each workload, \system is consistently the most competitive
system, outperforming the second-best alternative
by up to 4.0$\times$.
\jpb{Although unlimited reads is a key feature of \system, the multitude of new techniques that 
comprise its design ensure that \system also stands out 
in workloads characterized by RO or update transactions that fit the HTM capacity.}

We believe that, by achieving these gains, \system repositions HTM as a key building block for 
the new generation of applications \jpb{in-memory DBMSs} to embrace the upcoming CXL-based PM technologies.
Our results also highlight the power of suspend/resume access tracking in HTM.
In a hardware/ software co-design perspective, we believe that our results constitute solid evidence
that the recent trend led by Intel of introducing instructions for suspending load tracking in its 
latest 
processors should be generalized to suspension of both load and store tracking, 
as well as be adopted by other microprocessor manufacturers. 
Further, we also discuss how a subset of \system's features can be adapted to 
today's off-the-shelf Intel HTMs.

The paper is organized as follows. 
\S\ref{sec:background} presents background and related work, as well
as a preliminary study that highlights the challenges 
that lie ahead.
\S\ref{sec:algorithm} describes the design of \system.
\S\ref{sec:evaluation} experimentally evaluates \system.  
Finally, \S\ref{sec:discussion} discusses our work and \S\ref{sec:conclusions} concludes it.

\section{Background and Related Work}
\label{sec:background}

\jpb{The advent of PM brought new opportunities for building DBMS/OLTP engines. New hybrid designs (DRAM+PM) have become very popular~\cite{wbl, nvmdatabase, foedus, zen, zenplus, falcon}. In general, these systems exploit the byte-addressability of the new memory technology to optimize the logging mechanism and reduce write redundancy.}
\jpb{For instance, Write-Behind Logging (WBL)~\cite{wbl} logs the locations of the database that have changed instead of how they were changed. Zen~\cite{zen, zenplus} gets rid of the persistent transaction logs by persisting the updated values directly to PM at commit time (creating a new version). A garbage collector is then responsible to free stale values from PM.}

This section provides background on HTM and surveys recent proposals to 
extend HTM with durability and with unlimited reads.
We start by describing existing HTM systems in \S\ref{sec:background:htm}.
Then, we survey the state of the art on improving hardware transactions with  durability (\S\ref{sec:background:durable}) and unlimited reads  (\S\ref{sec:background:unlimitedreads}).
Finally, in \S\ref{sec:naive} we show that, although techniques from the two previous sections can be 
directly combined, such combination yields disappointing performance. 

\subsection{Hardware transactional memory}
\label{sec:background:htm}

HTM provides a hardware-based implementation of the familiar abstraction of atomic transactions. As of today, HTM implementations are offered by major CPU manufacturers such as Intel~\cite{yoo13sc}, ARM~\cite{ARMHTM} and IBM~\cite{le15ibm}. 
Despite their differences, all commercially available HTM implementations keep track of the transactions' accesses within per-core caches. 
%
This implies two fundamental limitations.

The \textbf{first limitation} is that, upon the commit of a transaction, its updates are not 
atomically flushed to PM.
\jpb{This is not an issue when hardware transactions exclusively access volatile objects 
since the cache coherence protocol ensures that the (cached) writes of committed transactions are immediately visible to other 
threads. 
However, } This becomes problematic if hardware transactions access persistent objects residing on a PM device. 
In fact, if the system crashes after a hardware transaction has committed, some of its writes on PM locations may not have yet persisted. Hence, they 
may be lost when the system later recovers. 

A \textbf{second limitation} stemming from the cache-centric design of current HTM systems is that current 
HTM systems abort \jpb{provide a \textit{best-effort} implementation of the transaction abstraction, in the sense that transactions are not guaranteed to commit even if they run in absence of concurrency. 
The main reason for that is} when a transaction's  footprint exceeds the transactional cache capacity. 
As such, 
applications must rely on an inefficient Single Global Lock (SGL) fallback mechanism. 


%

\sloppy Besides the basic primitives for transaction demarcation (\texttt{htmBegin}, \texttt{htmCommit}, \texttt{htmAbort}) some HTM implementations provide extensions
that let the program temporarily suspend the tracking of a certain kind of memory accesses, and later 
resume it.
%
During a suspend-resume window, the \emph{untracked} memory accesses are not added to the 
local transactional cache, hence they neither take up the capacity of such cache, nor are they considered for conflict detection. 
Evidently, inadvertently suspending access tracking can  lead to isolation anomalies. 
\jpb{It should also be noted that untracked accesses 
can still make other transactions abort --- by triggering the invalidation of the transactional cache of a concurrent transaction and its consequent abort.}





Among today's landscape of commercial HTM implementations, IBM POWER CPUs provide the richest set of primitives. 
The program can, at any point in a transaction's lifetime, suspend tracking of \emph{any access} and later resume.
Furthermore, tracking suspension of loads (but not stores) can be requested for the whole duration of a transaction by passing a specific flag to \texttt{htmBegin}. 
Transactions of this kind are also called \emph{Rollback-Only Transactions} (ROTs) \cite{le15ibm}.


As for Intel, 
the latest version of its Transactional Synchronization Extensions (TSX)
\jpb{, 
which equips the recent Intel Sapphire Rapids CPUs, }
has introduced the possibility of suspending load tracking
(with the new \texttt{XSUSLDTRK} and \texttt{XRESLDTRK} instructions \cite{intelManual}). 
On the other extreme of the spectrum, 
ARM's Transactional Memory Extension (TME) does not currently allow any control over access tracking.

Our work exploits access tracking suspension primitives in novel ways.
From here on, 
we consider the richest set of primitives, as offered by IBM POWER.
Later on, in \S\ref{sec:discussion}, we also discuss how our proposal can be adapted to Intel's HTM.

\subsection{Durable hardware transactions}
\label{sec:background:durable}


Expanding the abstraction of TM to interoperate with PM requires additional mechanisms. 
The usual requirement is to 
ensure that opacity~\cite{opacity}  (the established correctness criteria in 
transactional memory literature)
is preserved even if the system crashes and recovers, which
is captured by the formal notion of durable opacity~\cite{durableopacity20}.

A traditional approach is to use Write Ahead Logging (WAL) techniques~\cite{KolliASPLOS2016}, in 
which a transaction logs all modifications to PM before actually modifying them. 
The first works on persistent TM were based on software (PSTM), 
namely Mnemosyne~\cite{VolosASPLOS2011} and NVHeaps~\cite{Coburn2011NVHeaps}.
These early solutions catalyzed many subsequent improved PSTM designs~\cite{KolliASPLOS2016,LiuASPLOS2017,lazypersistency18,krish2020asplos,KJI+21,XIS21,SpecPMT_ASPLOS23,BaldassinCSUR2021}.
Among recent proposals, Pisces~\cite{Gu2019ATC} is the most related to our paper, since its design 
also prioritizes the performance of RO transactions. Pisces achieves this by leveraging the weaker semantics of Snapshot Isolation (SI) \cite{Berenson1995SIGMOD,Cerone2016PODC,Fekete2015}. 


Let us now focus on HTM. 
%
Since HTMs rely on cache coherency protocols and internal hardware buffers to handle  synchronization, 
one cannot repurpose such mechanisms to use WAL without modifying their implementation~\cite{avni2015, avni2016,wang15cal, atom, gileshtpm, morlog, joshi18isca}.
The reason is that 
existing HTM implementations disallow flushing caches from within an active 
hardware transaction -- the idea at the basis of WAL. \jpb{, which were originally designed to interoperate 
with software-based implementations of the transaction abstraction. }

The first systems to reconcile off-the-shelf 
HTMs with durability guarantees were proposed 
by cc-HTM~\cite{giles2017ISMM}, DudeTM~\cite{LiuASPLOS2017}, NV-HTM~\cite{CASTRO201963} and Crafty~\cite{GencPLDI2020}. 
More recently, SPHT~\cite{SPHT} revised 
such designs
to achieve enhanced scalability. \jpb{both during transaction processing 
and recovery. }
%
All these proposals (except for Crafty) share a common design backbone.
They rely on the OS's Copy-on-Write (CoW) support to 
transparently create a volatile working snapshot. 
Transactions operate on such a snapshot.
Transactional writes to the volatile snapshot are described in durable 
per-thread redo logs maintained in PM. Such redo logs are replayed on the 
original persistent heap in PM, typically in background, in order to bound the log's growth. 
Upon recovery, the durable redo logs are also replayed to reconstruct 
the durable contents of the persistent heap.

\begin{figure}[t] 
\centering
\includegraphics[width=0.45\textwidth]{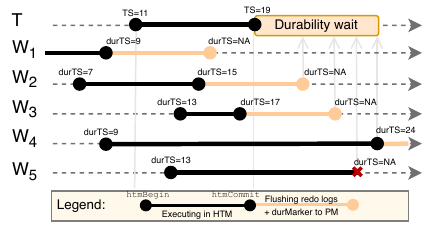}
 \vspace{-2mm}
\caption{Example execution with SPHT.}
\label{fig:RO-wait}
\end{figure}

The life cycle of a transaction is illustrated in Figure \ref{fig:RO-wait}. 
It comprises 
two phases. 
\jpb{, 
outlined in Figure \ref{fig:spht-outline} (left).}
First, the \emph{non-durable phase} occurs when a transaction, $T$, executes in HTM.
\jpb{(upon issuing the \texttt{htm\_begin} instruction).}
During execution, for each write (performed on the volatile snapshot), \jpb{DRAM-hosted shadow copy of the persistent heap,} an entry describing the write is added
to a per-thread redo log in PM.
%
Once $T$ has performed its last access, 
$T$ obtains a physical timestamp, also known as \emph{durability timestamp (durTS)}, before committing in HTM. 
This timestamp determines the order in which $T$'s redo log 
will be replayed, relatively to other transactions' redo logs. 
Since \emph{durTS} is obtained from within the hardware transaction, 
it is guaranteed that the \emph{durTS} order is consistent 
with the transaction serialization order determined by the HTM~\cite{CASTRO201963}.

Each thread's \emph{durTS} is a shared variable that can be observed by other threads.
Since assigning a physical clock to \emph{durTS} from within the 
hardware transaction may lead to spurious aborts, recent PHT solutions adopt the following 
optimization. Before committing in HTM, $T$ reads the physical clock to a private variable. 
Only after $T$ has committed in HTM does $T$ advertise the private timestamp in $T$'s \emph{durTS}.
For the sake of correctness (as we discuss next), a committed transaction cannot hold a null \emph{durTS}.
Therefore, each thread that is about to begin a transaction starts
by assigning an initial, conservatively low value to its \emph{durTS}. 

After executing, $T$ commits in HTM and, if successful, enters the \emph{durability phase}, 
which ensures that its writes are durable -- i.e., will be visible upon recovery if the system crashes   -- before the thread returns to the application.
%
Firstly, $T$ flushes each redo log entry. 
\jpb{However, at this point, $T$ cannot be considered durable yet. 
In fact, there may exist another transaction $T'$ whose effects $T$ may have observed
---thus, $T'$ must be replayed before $T$---,
but whose redo logs are not durable yet.}
Each transaction is only considered durable after
it has persisted a \emph{durable} marker (\emph{durMarker}) 
that includes \emph{durTS}. 
However, flushing the \emph{durMarker} to PM requires careful coordination with other threads, 
in order to ensure that an update transaction $T$ does not become durable (i.e., flushes its \emph{durMarker})
before any update transaction from which $T$ has read.



A possible approach to ensure the above property\jpb{(used, e.g., by NV-HTM~\cite{CASTRO201963}),} is to have $T$ first carry out a \emph{durability wait}, 
in which a transaction waits until every transaction with a lower \emph{durTS} has either become durable or aborted. 
Only then can the transaction flush its \emph{durMarker} to PM and return to the application. 

The durability wait is a well-studied bottleneck \cite{SPHT} and
Figure \ref{fig:RO-wait} illustrates why. Let us consider the durability
wait conducted by transaction $T$. 
Since $W_1$, $W_2$ and $W_3$
committed before $T$, $T$ may have observed  their writes. Hence, $T$ must wait until these transactions become durable. 
Still, there are less intuitive scenarios that force $T$ 
to wait also for transactions whose writes it never observes. 
This is the case with transactions $W_4$ and $W_5$:
$W_4$ will commit after $T$, whereas $W_5$ will simply abort. 
However, when $T$ checks their \emph{durTS}, it sees a lower value, which was conservatively 
set before they began. 
Therefore, $T$ must (spuriously) wait until their \emph{durTS} are updated (either upon commit or abort). 


Since RO transactions produce no persistent effects, 
they skip the redo log and \emph{durMarker} flush logic. 
However, they still have to go through the durability wait. 
Otherwise, a RO transaction could externalize (to the application program) 
the effects of transactions that are not yet durable.
As Figure~\ref{fig:motivation} has shown, the RO durability wait represents a strong performance 
penalty that, to our knowledge, no prior work in PHT has addressed.

\subsection{Unlimited reads with HTM}
\label{sec:background:unlimitedreads}


Recent proposals extend unmodified HTM implementations with a software layer that is 
able to stretch or even eliminate the \emph{read} capacity limitations of HTM~\cite{Felber2016EuroSys, Issa2018Middleware, IssaDISC2017, FilipePPoPP2019}. 
To our knowledge, these proposals focus on volatile hardware transactions (not PHT).
%
%
SI-HTM~\cite{FilipePPoPP2019} currently stands out as the only
solution that is able 
(i) to grant unlimited reads to both RO and update (hardware) transactions, 
while (ii) enabling multiple update transactions to run concurrently.
%
SI-HTM relies on the 
primitives for 
suspending access tracking.
In SI-HTM, update transactions are executed entirely with no load tracking, 
which 
grants them unlimited reads.
RO transactions run outside any transactional context, 
thus also benefit from unlimited reads, as well as no HTM overhead. 

\begin{figure}[t] 
\centering
\includegraphics[width=0.49\textwidth]{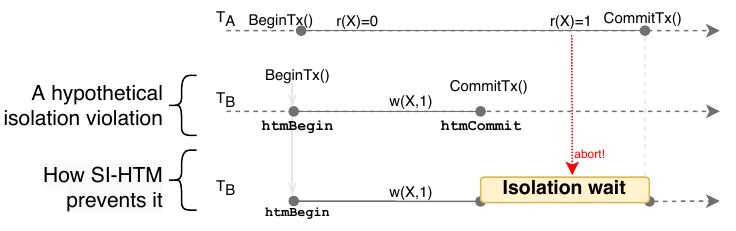}
\vspace{-1mm}
\caption{How SI-HTM prevents isolation violations.}
\label{fig:SI-violations-examples}
\end{figure}

This raises a consistency challenge: 
concurrent transactions with untracked loads may lead to isolation anomalies.
In the example in Figure~\ref{fig:SI-violations-examples}, 
transaction $T_A$ reads twice from the same memory location, $X$, whereas transaction $T_B$ concurrently 
modifies $X$ and commits. 
Since $T_A$'s first read happened before the concurrent write and the second read was after the write became 
committed, $T_A$ observes two different values from the same location, which 
constitutes an undesirable \emph{non-repeatable read} anomaly \cite{Berenson1995SIGMOD}.
(Note that, if these  reads were tracked by the HTM, it would abort $T_A$, thus preventing the anomaly.)
\jpb{Suppose that an update transaction, $T_B$, wishes to commit, while a
transaction, $T_A$, is running concurrently and has already performed some 
\emph{untracked} reads.
Hypothetically, if $T_B$ immediately committed its encompassing hardware transaction, this
could lead to a scenario where $T_A$, by performing a subsequent read, can observe
two different snapshots, thus, violating SI.}

SI-HTM prevents this anomaly by making update transactions that reach \texttt{CommitTx()} 
wait until every concurrent transaction that is active at that point is no longer active; i.e., has performed its last
transactional access, or aborted.
We call this an \emph{isolation wait}. Only after completing the isolation wait can an update transaction 
commit in HTM.


The bottom of Figure~\ref{fig:SI-violations-examples} illustrates how the isolation wait prevents the 
isolation anomaly in the previous example.
The isolation wait ensures that, if a waiting update
transaction (e.g., $T_B$) has performed any write that
is subsequently read by a concurrent transaction (e.g., $T_A$), 
a read-after-write conflict will occur while the former is waiting. 
The HTM detects the conflict (since the write is tracked) and aborts one of the conflicting transactions\footnote{This holds in all contemporary HTM implementations that support access tracking suspension (IBM POWER and Intel).}.
If the reader is a RO transaction (thus, running outside HTM), then the writer is always the victim, since
a RO transaction does not run in HTM.

\jpb{Figure~\ref{fig:spht-outline} (right) summarizes how SI-HTM puts the above recipe into practice. }
To put the above recipe into practice, each thread in SI-HTM shares a variable that advertises 
 the thread's current state, which can be:  \emph{inactive}, not executing a transaction; \emph{active}, executing a transaction; or \emph{waiting}, 
 performed the last memory access but still waiting for safety. 
For update transactions, the transition from \emph{active} to \emph{waiting} happens 
before the transaction has committed in HTM. 
To ensure that such state transition is immediately observable by other threads 
(which might also be in their isolation wait), the thread first suspends all access tracking, the new (\textit{waiting}) state of the transaction is externalized 
and access tracking is then resumed again. 
Finally, the thread scans the state of every other thread and spins until any thread initially found in the active state
has transitioned to a different state. 
As soon as that condition is met, the isolation wait is over and the thread can  commit in HTM.

It can be proven \cite{FilipePPoPP2019} that this isolation wait guarantees  
a fundamental property:

\begin{property}
    \label{prop:isolationwait}
    Given any pair of concurrent transactions\footnote{I.e., the time intervals between their invocations of \emph{beginTx} and \emph{commitTx} overlap.}, neither one can read from the writes performed by the other. 
\end{property}

Filipe et al. \cite{FilipePPoPP2019} show that this property is sufficient to 
provide SI.
However, it does not ensure that concurrent update transactions are opaque 
\cite{opacity} among each other.
A simple workaround to achieve opacity 
is to run update transactions as full HTM transactions (i.e., 
tracking loads). A drawback is that update transactions loose unlimited reads --  only RO transactions retain that advantage.



\subsection{Why not just combining durability with unlimited reads in PHT?}
\label{sec:naive}


\begin{figure}[t]
	\centering
	
	\begin{tabular}{ccc}
	\hspace{-2mm}\includegraphics[width=0.157\textwidth]
	{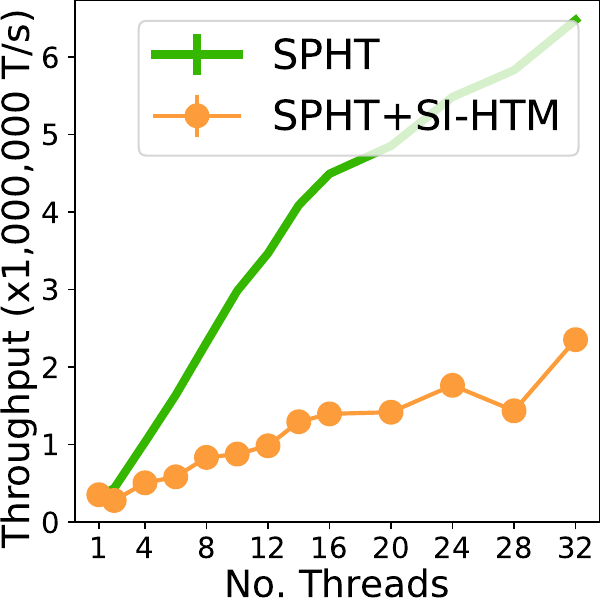}
	&
	\hspace{-2mm}\includegraphics[width=0.157\textwidth]
	{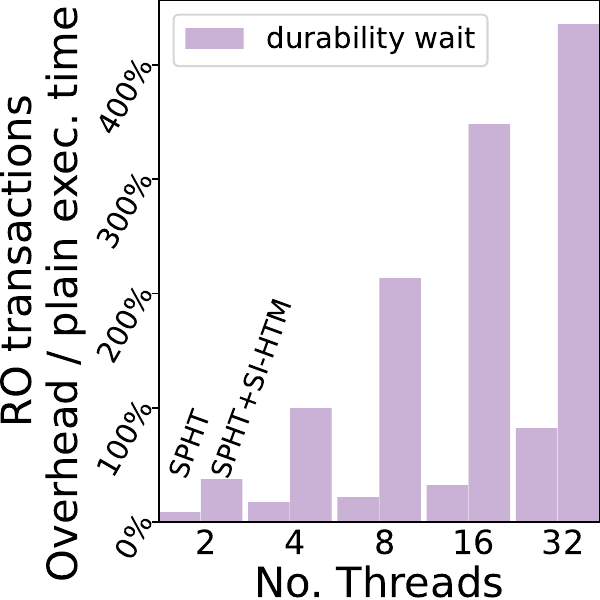}
	& \hspace{-3mm}\includegraphics[width=0.157\textwidth]{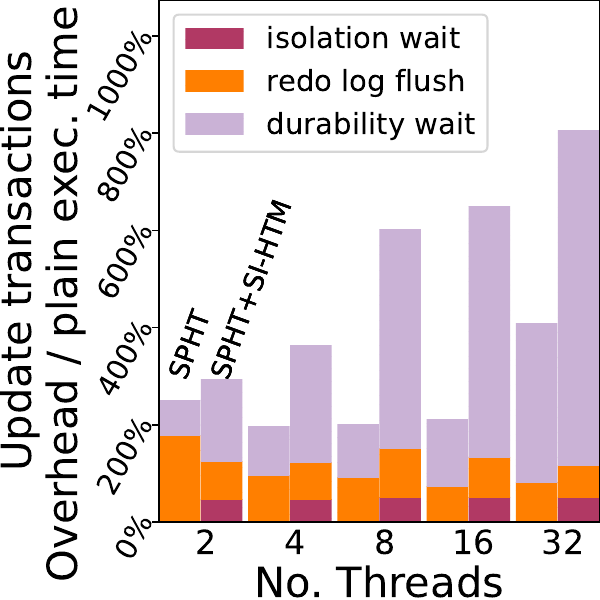}
	\\
	\end{tabular}
        \vspace{-3mm}
	\caption{Comparing the throughput of SPHT and SPHT+SI-HTM, in a mix of 95\% \emph{order status} and 5\% \emph{payment}
	transactions from TPC-C, with disjoint warehouse accesses} 
	\label{fig:naiveresults}
\end{figure}

To the best of our knowledge, no prior work has 
unified durability and unlimited reads in hardware 
transactions. 
Yet, such unification is
relatively straightforward to achieve. 
We start by adopting SPHT's architecture and design.
Further, we inject the main features of SI-HTM into update transactions, namely: 
run without load tracking and perform SI-HTM's isolation wait (before \texttt{htmCommit}).
Finally,  we replace the \texttt{htmBegin}/\texttt{htmCommit} instructions from RO transactions by 
routines that implement the \emph{active}/\emph{inactive} state transitions.


As a first exploratory step, we have implemented this no-frills SPHT+SI-HTM combination
and experimentally compared its performance with SPHT.
Unfortunately, this revealed a prohibitive side-effect of SPHT+SI-HTM. 

To understand it, Figure~\ref{fig:naiveresults} presents results 
obtained when running both systems with a mix of 95\% \emph{order status} RO transactions 
and 5\% \emph{payment} update transactions.
We disabled SMT and restrict each thread to access a disjoint warehouse. The former setting grants
large-enough transactional caches to prevent capacity aborts. The latter ensures negligible aborts due to transactional conflicts.
Hence, we can compare the durability overheads without noise incurred by different capacity/conflict abort rates of each solution. 

The left plot (in Figure~\ref{fig:naiveresults}) shows that
SPHT+SI-HTM exhibits considerably lower throughput than SPHT.
This disadvantage can be explained by two decisive differences that stand out when we
analyze the latency profile of each type of transaction.
First, the middle plot shows that RO transactions spend
up to 7$\times$ more time in their durability wait in SPHT+SI-HTM (than in SPHT).
\jpb{which constitutes a high price that RO transactions pay for unlimited reads. }
This is explained by a cascade effect:
by injecting an isolation wait in update transactions, they take longer to commit in HTM
(see the right plot); consequently, the durability wait conducted by RO transactions
also takes longer to complete.


Moreover, update transactions are also strongly penalized, as the right plot shows.
Not only do they need to include the isolation wait in their critical path, but 
also the post-HTM durability routines of SPHT now take considerably longer -- for the same reason as above.


%

\section{\system}
\label{sec:algorithm}

The previous preliminary study shows that, to effectively address the evident performance bottlenecks
that state-of-the-art PHTs inflict on RO transactions (recall \S\ref{sec:intro}), 
we need much more than a no-frills combination of techniques for durability and unlimited reads.
To accomplish this main objective, the design of \system combines some core mechanisms of SPHT and SI-HTM, 
and deeply rebuilds them by employing innovative techniques. 






\begin{figure}[t] 
    \centering
    \includegraphics[width=0.49\textwidth]{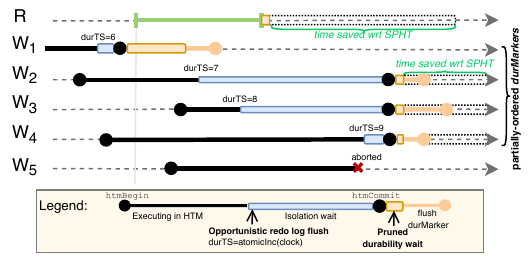}
     \vspace{-3mm}
    \caption{An example execution of \system, highlighting its main optimizations}
    \label{fig:psi-diagram}
    \end{figure}

\system adopts the base architecture of the state of the art in durable HTM 
(\S\ref{sec:background:durable}), in which transactions execute on a DRAM-hosted shadow copy of the persistent heap. 
\system comprises two main processes: the \emph{working process} and the \emph{log replayer}. 
The working process has parallel working threads that are 
responsible for executing the application (transactional and non-transactional) code.
When created, the working process receives an argument that specifies which isolation criterion it 
must enforce. \system supports both SI  
and opacity. 
This choice entails a trade-off:
with SI, every transaction benefits from unlimited reads; whereas, with opacity, only RO transactions do.

Transactions can access persistent locations that are originally
stored in a data file residing in PM.
The working process maps this PM file as a persistent heap in CoW mode (private mode in \texttt{mmap}). 
This mechanism transparently duplicates in DRAM any page that is altered.  
The working process also maps an additional persistent heap in shared mode, 
where it allocates the redo logs. Updates to this persistent heap are directly propagated to the  PM, bypassing the OS page cache.

The log replayer  process  replays the durable logs produced by transactions into the persistent heap. 
It is activated upon recovery, but can also be executed during transaction processing either in background (to prune the logs) or synchronously (when the logs' capacity is exhausted).

\begin{algorithm}[t]
\scriptsize
\tiny
\caption{\system} 
\label{alg:PSI2}
\begin{algorithmic}[1]
    \State \texttt{durMarkerArray}, \texttt{redoLog}[\text{N}] \CommentCL{Shared persistent variables (N is number of threads)}
    \State \texttt{volatileRedoLog}[\text{N}], \texttt{state}[\text{N}], \texttt{durTS}[\text{N}] \CommentCL{Shared volatile variables}


    \State \texttt{isReadOnly}, \texttt{beginTime} \CommentCL{Thread local volatile variables}

    \vspace{0.1cm}
    \Function{BeginTx(isolationLevel)}{} \label{alg:begintx}
    \State \texttt{beginTime} $\leftarrow$ getTime()
    \State $\texttt{state}[myTid] \leftarrow $ {\sc \{active, \texttt{beginTime}\}} \label{alg:begintx_set_active}
    \State $\texttt{durTS}[myTid] \leftarrow $ $-1$
    \State $\texttt{isReadOnly} \leftarrow false$
    \State {\sc memfence}
    \If {isolationLevel=SI}
    \State {\sc htmBegin(noLoadTracking)}
    \Else
    \State {\sc htmBegin(trackAnyAccess)}
    \EndIf
    \EndFunction

    \vspace{0.1cm}
    \Function{BeginTxReadOnly()}{} 
    \State \texttt{beginTime} $\leftarrow$ getTime()
    \State $\texttt{state}[myTid] \leftarrow $ {\sc \{active, \texttt{beginTime}\}}
    \State $\texttt{isReadOnly} \leftarrow true$
    \State {\sc memfence}
    \EndFunction
    
    \vspace{0.1cm}
    \Function{write}{addr, val} \label{alg:SPHT:write_inst}
    \State {\sc logWrite}(addr, val) \CommentCL{add redo log entry (without flushing it)}
    \State $*$addr$\leftarrow$ val  \CommentCL{execute write}
    \EndFunction
    
    \vspace{0.1cm}
    \Function{CommitTx()}{}\label{alg:commitTx}
    \If{{\sf isReadOnly}}
    \State $\texttt{state}[myTid] \leftarrow $ {\sc \{inactive\}}\label{alg:ro_set_inactive}
    \State {\sc DurabilityWait()}
    \Else
    \State {\sc htmSuspendAnyAccessTracking}\label{alg:sus}
    \State $\texttt{state}[myTid] \leftarrow $ {\sc \{inactive\}}\label{alg:inactive}
    \State {\sc memfence}
    {\State {\sc flushRedoLog(volatileRedoLog[myTid], redoLog[myTid])}\CommentCL{Flush redo log opportunistically}}\label{alg:optredoflush}
    \State $\texttt{durTS}[myTid]\leftarrow $ $\textsc{atomicInc}(\texttt{globalOrderTs})$ \CommentCL{Acquire the next durTS}\label{alg:atomicinc}
    {\State {\sc IsolationWait()}}\label{alg:isowaitcall}\CommentCL{Similar to SI-HTM}
    \State $\texttt{state}[myTid] \leftarrow $ {\sc \{non-durable,getTime()\}}\label{alg:nondurable}
    \State {\sc htmResumeAccessTracking}
    \State {\sc htmCommit}
    \State {\sc memfence}\label{alg:memfenceBeforeDurWait}
    \State {\sc DurabilityWait()}
    \State {\sc FlushDurMarker}(redoLog[myTid], $\texttt{durTS}{[myTid]}$, durMarkerArray)\label{alg:FlushDurMarker}
    \State $\texttt{state}[myTid] \leftarrow $ {\sc \{inactive\}}\label{alg:aborMarker}
    \EndIf
    
    \EndFunction
    
    \vspace{0.1cm}
    \Function{IsolationWait()}{}\label{alg:isowait}
    \State $\texttt{state\_snapshot}[0 .. \text{N-1}] \leftarrow $ $\texttt{state}[0.. \text{N-1}] $
    \For{$c \in [0 .. \text{N-1}]: c \ne myTid$}
    \If{state\_snapshot[$c$].isActive}
    \State \textbf{wait while }{$\texttt{state}[c] = \text{state\_snapshot}[c]$}
    \EndIf
    \EndFor
    \EndFunction
    
    \vspace{0.1cm}
    \Function{DurabilityWait()}{}\label{alg:durwait}\CommentCL{Pruned durability wait}
    \State $\texttt{state\_snapshot}[0 .. \text{N-1}] \leftarrow $ $\texttt{state}[0.. \text{N-1}] $
    \For{$c \in [0 .. \text{N-1}]: c \ne myTid$}
    \If{state\_snapshot[$c$].isNonDurable and state\_snapshot[c].time $<$ \texttt{beginTime}} \label{alg:durwaitif} 
    \State \textbf{wait while }{$\texttt{state}[c] = \text{state\_snapshot}[c]$}
    \EndIf
    \EndFor
    \EndFunction

    \vspace{0.1cm}
    \Function{abortHandler()}{}
    \State $\texttt{state\_ts}[myTid] \leftarrow $ {\sc [INACTIVE]}
    \If{$\texttt{order\_ts}[myTid] \ne -1$} \label{orderts-1}
    \State {\sc writeAbortMarker()} \CommentCL{Writes abort marker (asynch.)}
    \EndIf
    \EndFunction

	\end{algorithmic}
\end{algorithm}

Figure \ref{fig:psi-diagram} showcases \system's working process in action, highlighting the main 
optimizations that it employs.
Its pseudo-code is reported in Algorithm \ref{alg:PSI2}.
In the 
following sections, we describe it in detail. 
We start with its non-durable phase (\S\ref{sec:nondurableexec}) 
and then proceed to the durability phase (\S\ref{sec:durphase}). 
Finally, we address \system's log replayer (\S\ref{sec:logreplay}).


\subsection{Non-durable execution phase}
\label{sec:nondurableexec}

We start by focusing on the \textit{non-durable execution phase}, which operates on the 
volatile snapshot of the persistent heap. 
\jpb{We start by focusing on the initial execution phase, which operates on the 
non-durable shadow copy of the persistent heap. 
As such, we refer to this phase as \textit{non-durable execution phase}.}
\jpb{The novelty of \system's non-durable execution phase (versus existing designs
for PHT) is that it provides unlimited reads to RO and/or update transactions.
The algorithm assumes that the underlying HTM offers full support to suspend/resume access (from loads to any accesses).
Currently, this is offered by the IBM POWER9 implementation. 
In \S\ref{sec:conclusions} we discuss how \system could be adapted to run on HTM implementations with
more restrictive suspend/resume support (such as the latest generations of Intel TSX).}
Each thread 
has a reserved entry in a 
shared \texttt{state} array (as in SI-HTM), 
which announces the thread's current state.
Besides the original  
\emph{inactive} and \emph{active} 
states, 
we add a \emph{non-durable} state, which we detail in the next section.
The \emph{active} and \emph{non-durable} states are coupled with a physical 
timestamp, obtained from the CPU clock just before the thread enters such states. 
The physical timestamps serve two purposes in \system, which we explain shortly.

When the application calls the \emph{BeginTx} routine, 
the thread starts by setting its \texttt{state} to \emph{active} (ln.\ref{alg:begintx_set_active}).
In the case of RO transactions, \emph{BeginTx} returns 
without starting any HTM transaction. 
For update transactions, a hardware transaction is
started and \emph{BeginTx} returns (as in SI-HTM). 
If \system is configured to enforce opacity, 
new hardware transaction is created as a regular HTM transaction.
Otherwise, 
the hardware transaction starts without load tracking (as in SI-HTM).

To write to a persistent location, 
the thread calls the \emph{write} routine, 
which adds a new entry to the thread's redo log, 
describing the updated address and value, 
before %
performing the write.
This is analogous to prior PHT proposals (\S\ref{sec:background:durable}).

When the application intends to commit a transaction, it
invokes the \emph{CommitTx} routine. 
In the case of a RO transaction, the thread simply changes its state 
to \emph{inactive} (ln.\ref{alg:ro_set_inactive}).
Otherwise, 
the thread takes the following steps to guarantee
Property \ref{prop:isolationwait}:
it  suspends any access tracking in the encompassing hardware transaction (ln.\ref{alg:sus}), 
then it announces the \emph{inactive} state (ln.\ref{alg:inactive}), before it 
calls the \emph{IsolationWait} routine (ln.\ref{alg:isowaitcall}).
This routine 
waits until every other transaction that was active at the beginning 
of the isolation wait is no longer in that state (as in SI-HTM). 
An important detail in this routine is that the state timestamp lets the 
waiting thread disambiguate between an \emph{active} 
thread that is still running a concurrent transaction (hence must be waited for), 
and the same thread after having completed the previous transaction and started a new
one (at a higher timestamp).
(For now, let us skip lines \ref{alg:optredoflush} and \ref{alg:atomicinc}, and explain 
them in \S\ref{sec:durphase}.)
Finally, the thread changes its state to \emph{non-durable} and 
resumes access tracking, before it finally commits in HTM.

\jpb{Notice that all state transitions must occur when the thread is either 
not running in HTM or has suspended any transactional access 
tracking. 
Otherwise, not only such transitions would not be visible to other threads, but 
also the transaction that set a new state would abort if other threads attempted to read from state.}
Whereas SI-HTM only had one state transition in the beginning of its isolation wait, 
\system requires an additional state transition just after the isolation wait's end.
Therefore, \system stretches the \emph{suspended} window to cover both transitions.
We have empirically confirmed that this strategy is more efficient than suspending/resuming twice.

\jpb{It is also worth noting that the state timestamp plays an important role in the liveness of the isolation wait, 
since it enables the waiting thread to disambiguate between an \emph{active} 
thread that is still running a concurrent transaction -- hence must be waited for --, 
and the same thread after having completed the previous transaction and started a new
one (at a higher timestamp).}

\subsection{Durability phase}
\label{sec:durphase}

As soon as a transaction becomes  
non-durably committed, it enters the \emph{durability phase}. \jpb{but they may not be durable yet, hence cannot 
immediately return to the application.}
Recalling \S\ref{sec:background:durable}, this is a potentially cumbersome phase: 
RO transactions
must undergo a potentially expensive durability wait;
while update transactions execute even more costly steps 
(flushing the redo log, running the durability wait, and flushing the \emph{durMarker}).

In \system, we rethink each durability step, by proposing three novel optimizations.
The first one, \emph{pruned RO durability wait}, is able to practically 
hide 
the length of the durability wait of RO transactions. 
By combining this optimization with the unlimited reads feature (as 
described in the previous section), \system is able to fulfill the main 
goal of our paper -- to address the two bottlenecks that
hinder RO transactions' performance in existing PHT systems (recall \S\ref{sec:intro}).

Since update transactions in \system are penalized 
with a new time-consuming coordination step   
(the isolation wait), we compensate it with two optimizations to their durability phase,
\emph{opportunistic redo log flushing} and \emph{partially-ordered durability markers}.
The next sections describe these optimizations, focusing on RO transactions first (\S\ref{sec:ro_dur_wait}), 
then update transactions (\S\ref{sec:redologflush} and \S\ref{sec:durmarkers}). 

\subsubsection{Pruned RO durability wait}
\label{sec:ro_dur_wait}

Recall from \S\ref{sec:background:durable} that
a RO transaction, $R$, cannot return to the application before 
every transaction from which $R$ read 
is durable.
In SPHT, 
$R$ conservatively 
waits for any update transaction that
had already committed before $R$ began, as well as any update transaction that executed 
concurrently with $R$.

Fortunately, we can take advantage of Property \ref{prop:isolationwait}.
A corollary of this property is that 
$R$ will never read from update transactions that executed concurrently with $R$. 
Therefore, we can safely \emph{prune} the RO durability wait to bypass any transactions that had not 
yet HTM-committed when $R$ began.
Concretely, in the \emph{DurabilityWait} routine,
a thread 
scans the state of all other threads and, whenever it finds
any thread in \emph{non-durable} state with a timestamp smaller than its own  
\emph{active} timestamp, it spin-waits until that state changes.%
\footnote{For simplicity, we assume perfectly synchronized hardware timestamp counters. As in prior research ~\cite{CASTRO201963}, this waiting step can be generalized to account for the maximum deviation offsets between any two clocks.}

This optimization has a twofold advantage, which 
 Figure~\ref{fig:psi-diagram} illustrates by revisiting the example from Figure~\ref{fig:RO-wait}.
As a first advantage, by
drastically reducing the number of transactions $R$ needs to wait for, it returns much earlier to the application.
In this example, 
$R$ only needs to wait for $W_1$ to become durable, 
since every other (concurrent) transaction is bypassed.
Furthermore,
since the pruned set only comprises transactions that have
committed earlier, it is likely that these are already durable by the time 
$R$ 
enters the durability wait.
This is the case with $W_1$ in our example. 


In RO-dominated workloads, most accesses to the \texttt{state} array are writes by RO transactions 
(as they change state).
These state transitions are read by the durability wait but have no effect on its logic. 
If left unaddressed, this interleaving of write/read accesses from different cores to the same cache lines 
would trigger many cache invalidations.
We prevent this thrashing condition by unfolding the per-thread state into two arrays 
(not presented in Algorithm \ref{alg:PSI2}, for simplicity).
The first array specifies whether a transaction is \emph{active} or not; the second array 
tells whether a transaction is in the \emph{non-durable} state or not. 
The \emph{DurabilityWait} routine scans the second array, which is only changed by update transactions.
Therefore, in RO-dominated workloads, cache invalidations on the second array become rare 
and \emph{DurabilityWait} serves most of its reads from L1 cache.



\subsubsection{Opportunistic log flushing}
\label{sec:redologflush}

Next, we focus on optimizations that reduce the overheads of the durability steps of update transactions
by exploiting the isolation wait. 
%
We start by 
focusing on the redo log flushing step. 

In every HTM implementation that we are aware of, 
hardware transactions cannot flush the cache lines they has written to 
(as that would externalize the writes). 
However, while a transaction is in (full) suspended mode, 
it is allowed to perform (untracked) writes and to write-back (without invalidating) their cache lines, 
as long as such cache lines had not been accessed by the transaction 
before it suspended access tracking.
Specifically, this is the behavior of IBM POWER.

\system's isolation wait, which is performed in \emph{suspended} mode, is 
an opportunity to exploit the above feature to 
\emph{opportunistically} anticipate the flushing of the cache lines of its redo log entries.
Two implementation details are crucial to accomplish this goal.
First, the redo log entries that are written by the \emph{write} routine 
are part of the transaction's write-set (in the transactional cache).
Therefore, they cannot be flushed in suspended mode.
To work around this restriction, each thread uses a volatile redo log (ln. 2) while 
executing transactionally and, in the \emph{suspended} window, copies each volatile redo log entry to 
a the persistent redo log, and only flushes the latter.
Second, 
the flush instructions are issued 
asynchronously.
Therefore, 
the thread can immediately proceed to the isolation wait while, in background, 
the CPU is flushing the cache lines to PM.
%
After committing in HTM, 
the thread executes a memory fence (ln.\ref{alg:memfenceBeforeDurWait}) to ensure 
 that any in-flight cache line flushes terminate before the thread 
continues to the next durability step.
Our results in \S\ref{sec:evaluation} show that, 
in most cases, such flushes have already completed
\emph{before} this point. 
When this happens, 
the latency of redo log flushing is effectively hidden behind the 
isolation wait.


\subsubsection{Partially-ordered durability markers} 
\label{sec:durmarkers}

The  two last steps of the durability phase of
update transactions in \system are notably simple. 
%
An update transaction performs the same pruned durability wait as RO transactions (see \S\ref{sec:ro_dur_wait}).
Then, it can immediately flush its \emph{durMarker} (ln.\ref{alg:FlushDurMarker}).

Hence, \system trades the total order among \emph{durMarker}s 
-- which, to our knowledge, underpins all prior PHT proposals -- for a
lightweight \emph{partial order}.
To illustrate, let us revisit the example in Figure \ref{fig:psi-diagram}.
The \emph{durMarker} of the concurrent transactions $\{W_2, W_3, W_4\}$ can be flushed 
in any order; they must only be ordered after $W_1$'s, 
This obviates any need for coordination across concurrent transactions 
(e.g., SPHT's intricate group commit scheme \cite{SPHT})
at this stage. 

The same advantages that we get by pruning the durability wait of RO transactions
(see \S\ref{sec:ro_dur_wait})
are also granted to update transactions. 
Moreover, the same correctness arguments that we provide in \S\ref{sec:ro_dur_wait} 
also apply to update transactions.
Concretely, once an update transaction, $T$, that has gone through its durability wait,
it is guaranteed that any transaction from which $T$ might have read is already durable,
so $T$ can also safely become durable. This is a corollary of Property \ref{prop:isolationwait}.
In fact, if a set of concurrent update transactions is performing their durability phase
and, due to a crash, only an arbitrary subset of them is durable upon recovery, 
we know that no write performed by the lost (concurrent) transactions can have been read by 
any durable transaction that survived.

\subsection{Efficiently replaying partially-ordered logs}
\label{sec:logreplay}


In prior PHTs, each logged transaction contains a \emph{durMarker}. 
Hence, the log replayer (LR) needs to scan through 
every per-thread's log to determine the next update transaction to replay.
This is a well-studied  bottleneck \cite{SPHT}.
To our knowledge, the only technique to avoid this scanning phase 
-- SPHT's log linking scheme \cite{SPHT} -- relies on totally-ordered \emph{durMarker}s.
Hence it is not compatible with \system.

To achieve a similar scalability to SPHT while supporting partially-ordered \emph{durMarker},
\system organizes every \emph{durMarker} in a global \emph{durMarker} array (implemented as a circular array).
The size of the array determines how many update transactions can be durable in the 
\emph{durMarker} array before having their writes replayed on the persistent heap. 

\system uses logical timestamps as \emph{durTS}, which serve as indexes to the \emph{durMarker} array.
As soon as a transaction, $T$, acquires its unique \emph{durTS}, $T$
becomes the owner of the corresponding entry in a global \emph{durMarker} array.
Each \emph{durMarker} entry in the array includes: the start address of $T$'s redo log, the number of entries, as well 
as $T$'s \emph{durTS}.

The \emph{durMarker} global array fulfills our scalability requirement, since the 
LR thread can just traverse the \emph{durMarker} array and
sequentially replay the redo log pointed to by each \emph{durMarker} entry that it reads.
After a batch of entries is replayed, the tail of the circular array is advanced, which
effectively frees such entries to new transactions. 


The choice of logical timestamps is fundamental to enable the solution described so far.
To ensure that the \emph{durTS} obtained by a transaction
is consistent with the isolation order determined by the HTM, the \emph{durTS} needs to 
be acquired before the transaction HTM-commits. 
The straightforward solution of reading and incrementing the global clock using transactional 
loads and stores is well-known to harm scalability under contention \cite{LiuASPLOS2017,SPHT}. 

To avoid this shortcoming, \system takes advantage of the
\emph{suspended} window of its isolation wait.
Concretely, the logical timestamp is acquired via an atomic 
increment instruction executed from 
such window (ln.\ref{alg:atomicinc}).
Since the atomic increment is not tracked by the HTM, it does not introduce transactional conflicts.
Moreover, similarly to opportunistic redo log flushing (\S\ref{sec:redologflush}), 
the latency of the atomic instruction is normally hidden as it is overlapped with
the (typically longer) isolation wait.

A transaction that has acquired a \emph{durTS} and then aborts will produce a \emph{hole}
in the logical timestamp sequence.
Flushing an \emph{abort marker} in the abort handler of such transactions 
to the corresponding entry in the global \emph{durMarker} array (ln.\ref{alg:aborMarker}) fixes the \emph{hole}. 
%
%
A system crash can also create holes in the \emph{durMarker} array. 
To illustrate this, let us recall the example in Figure \ref{fig:psi-diagram}, where
transactions $W_3$ to $W_5$ run concurrently and commit in HTM.
Transaction $W_4$ took longer than $W_3$ and $W_5$ to 
flush its \emph{durMarker} (this is possible due to \system's partially-ordered \emph{durMarker}s). 
Assume that a crash happens just before $W_4$ persists
its \emph{durMarker}. 
Hence, upon recovery, the array will hold a \emph{durMarker} in the entries 
corresponding to  the \emph{durTS} of $W_3$ and $W_5$, and a hole between them.
A crash-induced hole can be detected since it either has a null entry or an entry with an expired \emph{durTS} 
(i.e., from a previous epoch of the circular array). Let us designate these as \emph{unmarked} holes.


When the LR detects a hole, it jumps to the next valid entry. 
It is easy to show that there cannot be more than $n-1$ (crash-induced) \emph{unmarked} holes 
before the last valid \emph{durMarker} in the array.
Therefore, 
as soon as the LR has found $n$ \emph{unmarked} holes, it is guaranteed that there are no more
valid entries to replay and the LR can stop. 

\section{Evaluation}
\label{sec:evaluation}

Our evaluation aims at answering the following questions: 

\begin{enumerate}
\item \emph{In RO workloads of different read footprints, what advantage do \system's unlimited reads
	provide?} (\S\ref{sec:eval:ro})

 \item    \emph{In update-only workloads, given the isolation wait penalty and the new optimizations to the durability phase of \system, 
	what is the net performance outcome of both factors}? 
 (\S\ref{sec:eval:upd})
    
\item \emph{In mixed workloads (combining RO and update transactions) do the main observations taken from 
	the previous scenarios still hold?} 
 (\S\ref{sec:eval:mixed})

\item \emph{What is the performance of \system's log replayer when compared to the alternative systems? (\S\ref{sec:eval:logreplay})}

 \end{enumerate}



\subsection{Experimental settings}

\noindent\textbf{System.} 
We used a dual-socket IBM POWER9 machine equipped with DD2.3 CPUs at 2.3-3.8GHz and 1023GB DRAM.
Each CPU has 16 cores with simultaneous multi-threading (SMT).
In this architecture, HTM can only be used from a virtual machine (VM) \cite{power9:suspendtrap}, 
so
we set up a QEMU/KVM VM with 64 virtual cores (mapped to the 32 physical cores, each with 2 SMT hardware threads) and 8GB DRAM.
The suspend (and resume) access tracking instructions trigger a trap to activate KVM hypervisor, which implements (in software) the logic associated with these instructions \cite{power9:suspendtrap}.
This incurs a considerable penalty on such instructions.
Executing both instructions in sequence takes from 350ns on a single thread, 
to 1500ns on 64 threads. 

Since PM is not yet available for IBM POWER systems, we emulate the write latency of an 
Optane-like PM device connected via CXL each time a cache flush is requested.
Using access latency figures from the literature \cite{10.1145/3492321.3519556, 10.1145/3582016.3582063},
we inject a spin loop of 310 ns on each cache line flush. 
This approach is analogous to previous works in the literature (e.g., \cite{LiuASPLOS2017,CastroIPDPS2018}), whenever PM is not available.

\noindent\textbf{Evaluated solutions.} 
We used a full implementation of both variants of \system (\system-opa and \system-SI).
We compare them against two systems that, 
to our knowledge, are the most competitive proposals in the domains of PHT, SPHT \cite{SPHT}; 
and read-optimized PSTM, Pisces \cite{Gu2019ATC}.
We used the full-fledged SPHT variant, with forward log linking enabled~\cite{SPHT}.
\jpb{The log linking feature enables superior log replay performance but degrades transaction processing performance.}
\jpb{For Pisces, since its source code was not available, we implemented its algorithm from scratch,
with two simplifications to the original design: 
instead of storing each \texttt{next} pointer next to the corresponding object, 
we maintain such pointers at a word-granularity in a lock table as in standard STMs (e.g., \cite{5438987}); further, we disable log reclamation.}
We also considered pure HTM (with the SGL fallback), just as a reference of the raw throughput of HTM. 

For fairness, we implemented all systems in a common framework. 
The log replayer is disabled during transaction processing for
all solutions that support asynchronous log replay (\system and SPHT), 
which is coherent with previous papers in this category \cite{SPHT, CastroIPDPS2018}. 
Every solution based on HTM falls back to SGL after 10 retries.
The results are an average of 3 runs, each taking 5 seconds.

\noindent\textbf{Workloads.} As benchmarks, we use a full implementation of the TPC-C OLTP benchmark~\cite{tpcc}.
It uses a B-tree implementation that 
is exempt from SI's consistency anomalies (e.g., write skews~\cite{Fekete2015,Cerone2016PODC}).
As Table \ref{tab:tpcc_summary} shows, the transaction types cover a diverse read and write footprints mix. 

\begin{table}[t]
	\scriptsize
\caption{Read and write footprints in TPC-C}
\vspace{-2mm}
    \label{tab:tpcc_summary}
	\begin{tabular}{|l|l|l|}
	\hline
	Transaction type & Read footprint & Write footprint \\ \hline
	stocklevel & Very high (avg 122K) & none \\
	orderstatus & Moderate (avg 650) & none \\
	delivery & Very high (avg 86K) & Moderate (avg 30) \\
	payment & Low (avg 97) & Low (avg 5) \\
	neworder &	High (avg 7.5K)	& Moderate (avg 141) \\\hline
	\end{tabular}
\end{table}


\begin{figure}[t]
	\centering
        $\quad\;\,$\includegraphics[width=0.4\textwidth]
	{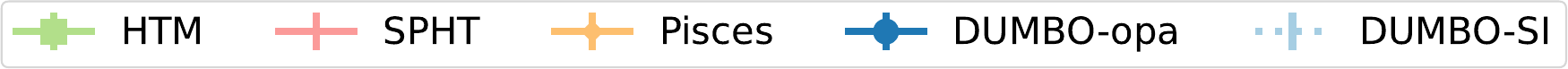}
        \begin{tabular}{c c}
        \hspace{-3mm}\includegraphics[width=0.25\textwidth]
	{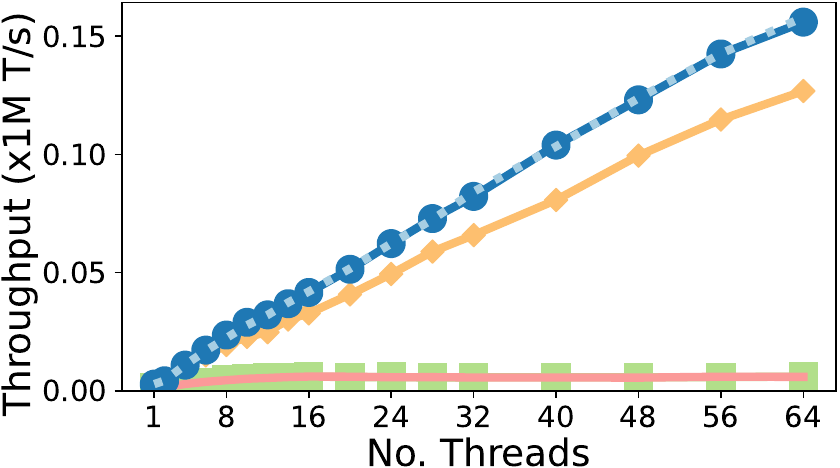}
	&
	\hspace{-3mm}\includegraphics[width=0.2325\textwidth]
	{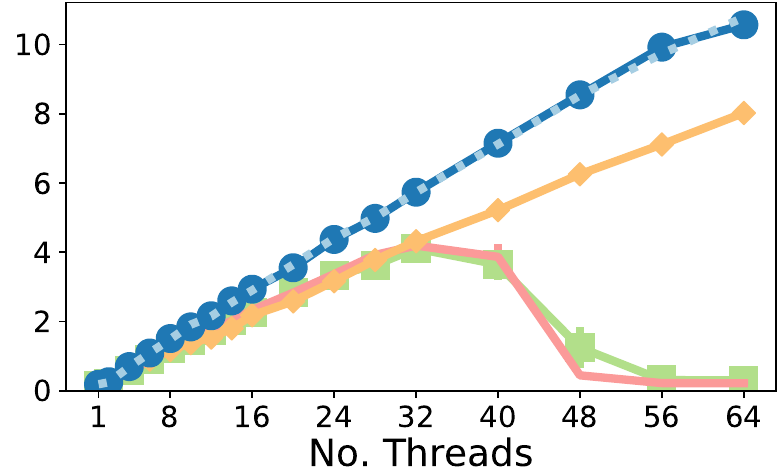}
	\\
	\end{tabular}
	\begin{tabular}{c c}
	\hspace{-3mm}\includegraphics[width=0.2485\textwidth]
	{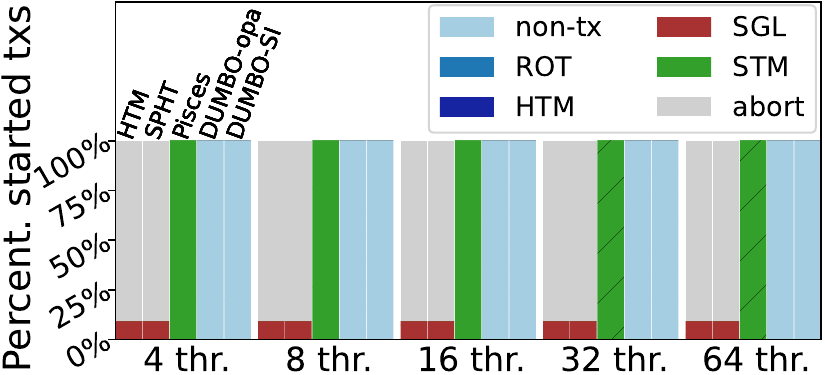}
	&
    \hspace{-3mm}\includegraphics[width=0.235\textwidth]
	{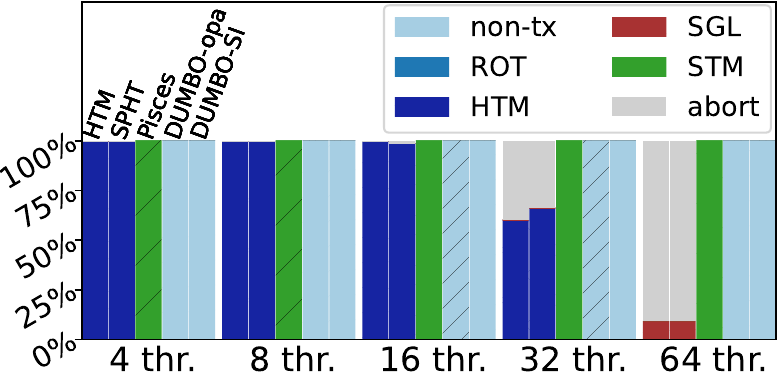}
	\\
	stocklevel & orderstatus \\
	\end{tabular}
  \vspace{-3mm}
	\caption{Throughput and transaction outcomes with RO TPC-C workloads}
	\label{fig:tpcc_readonly}
\end{figure}

\subsection{Processing RO workloads}
\label{sec:eval:ro}

Figure \ref{fig:tpcc_readonly} presents the throughput and transaction outcomes when exclusively 
running  \emph{stocklevel} and \emph{orderstatus} transactions. 
In both, the performance of \system stand out when compared to their persistent competitors.

Since \emph{stocklevel} transactions have large read footprints, 
they always capacity-abort when 
executed inside full HTM hardware transactions. This holds for SPHT as well as HTM. 
\jpb{In the presence of repeated capacity aborts, these systems always resort to the SGL fallback to 
execute transactions, which explains why they achieve no speed-up.}
In contrast, the remaining systems 
scale up to 64 threads. 
In \system, 
this is an expected consequence of the fact that RO transactions run without HTM. 
\jpb{, hence with unlimited read capacity.}
As in most STM implementations, Pisces imposes no limits on the read (or write) footprints.
While \system does not add any instrumentation to read accesses, Pisces instruments each read to 
a shared location in PM with a software routine that checks the lock table to determine the latest version to read from.
This explains the 28\% performance gap between \system and Pisces.

In the \emph{orderstatus} workload, the read footprints fit a full per-core transactional cache.
Therefore, as long each thread runs in its own core (up to 32 threads), capacity aborts are negligible, thus
the unlimited reads feature of \system 
has no benefit
when compared to SPHT/HTM.
Yet, we can still observe a performance advantage of $38\%$ 
due to 
the fact that SPHT/HTM need to issue \texttt{htmBegin}/\texttt{htmCommit} to activate the HTM support.
Beyond 32 cores, SMT colocates pairs of threads on a single core, hence the per-thread transactional caches no longer
suffice for most transactions and we see the same trends of the \emph{stocklevel} workload.

\subsection{Processing update-only workloads}
\label{sec:eval:upd}

\begin{figure}[t]
	\centering
        $\qquad$\includegraphics[width=0.4\textwidth]{ThroughputLegend_Hor.pdf}
        \begin{tabular}{c c}
	\hspace{-2mm}\includegraphics[width=0.247\textwidth]
	{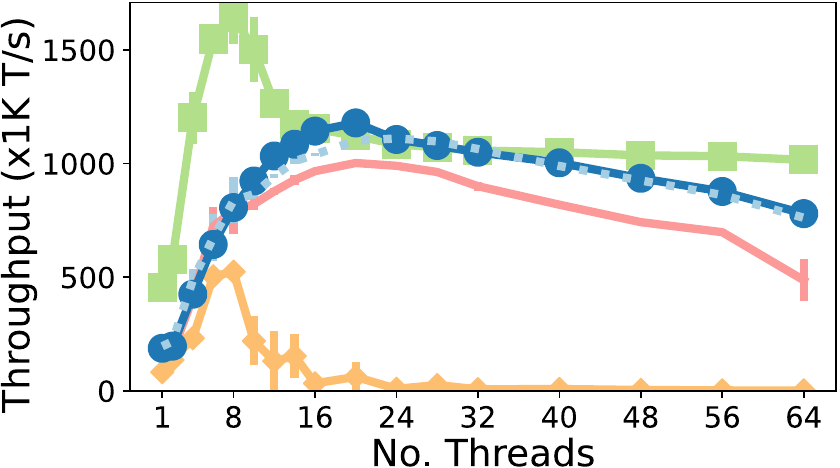}
	&
	\hspace{-3mm}\includegraphics[width=0.22\textwidth]
	{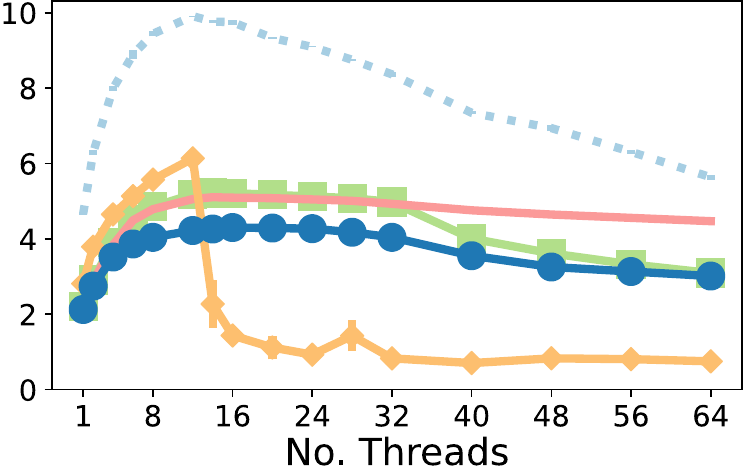}
        \\
	\hspace{-2mm}\includegraphics[width=0.24\textwidth]
	{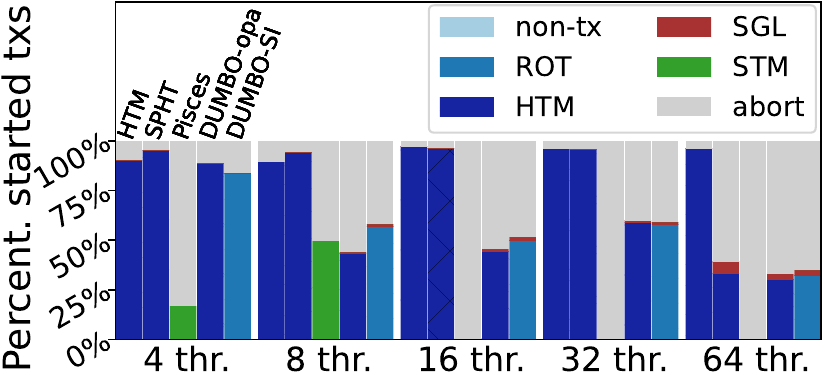}
	&
        \hspace{-3mm}\includegraphics[width=0.23\textwidth]
	{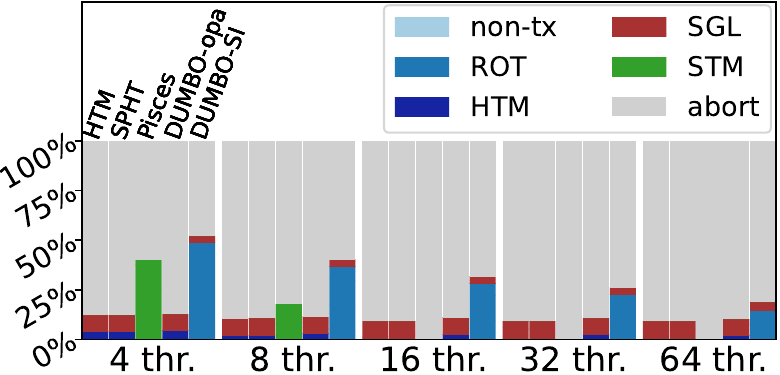}
	\\
	\hspace{-2mm}\includegraphics[width=0.24\textwidth]
	{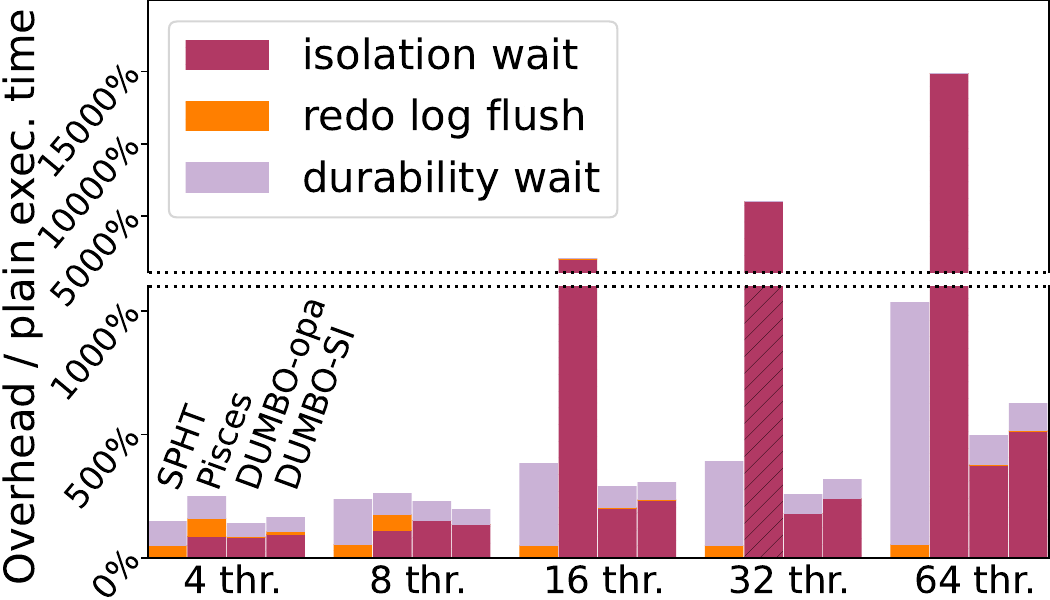}
	&
	\hspace{-2mm}\includegraphics[width=0.22\textwidth]
	{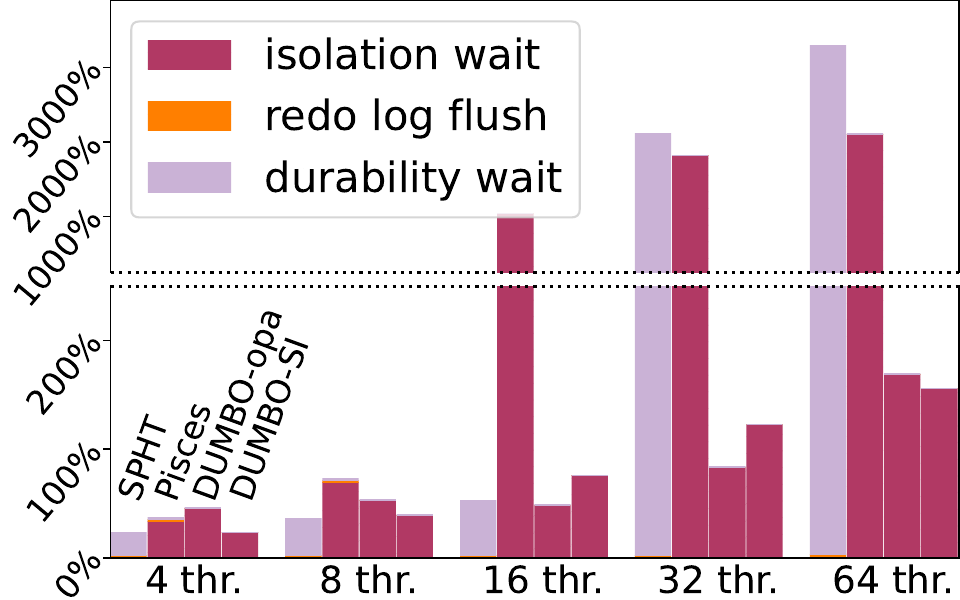}
	\\
	payment & delivery \\
	\end{tabular}
  \vspace{-3mm}
	\caption{Throughput, transaction outcomes and overhead breakdown with update-only TPC-C workloads}
	\label{fig:tpcc_updonly}
\end{figure}

\jpb{We now switch our attention to workloads exclusively comprising update transactions.}
Figure \ref{fig:tpcc_updonly} presents the results for the update-only \emph{payment} and \emph{delivery} workloads.
In contrast to the previous section, we now include a third graph (bottom of Figure \ref{fig:tpcc_updonly}),
which reports the overhead that committed transactions incur (on average) on the different steps
that occur beyond the plain execution (i.e., the interval between the invocations to \emph{BeginTx} and \emph{CommitTx}), 
relatively to time spent executing the latter. 
For instance, an overhead of 200\% for a given step means 
committed transactions spend $2\times$ more time performing that step than their plain execution.

We start by considering the \emph{payment} workload, which has a small footprint and negligible capacity aborts.
Hence, the \emph{unlimited reads} feature of \system has no impact in this workload, 
whereas the isolation wait and the durability optimizations of \system become decisive. 

We start by comparing \system with SPHT. \system-opa and \system-SI outperform SPHT by 11\% and 17\%, respectively. 
Looking at the bottom plot in Figure \ref{fig:tpcc_updonly}, we can see that, 
in SPHT, the overhead of the durability steps increases steeply after 8 threads, growing 
up to 1000\% (i.e., $\approx10\times$ the plain execution time) at 64 threads. The durability wait is the major contributor to such overhead.

In contrast, in \system, the main source of overhead is the isolation wait, 
whereas the durability overheads are substantially reduced (up to 2 orders of magnitude lower than SPHT's).
This reduction stems from the two optimizations in \system's durability phase. 
First, \emph{opportunistic redo log flushing} reduces the critical-path waiting time
to negligible (at peak throughput, 
0.93\% overhead in \system vs. 54\% in SPHT).
\jpb{The reason is that most of the PM write latency is overlapped with the isolation wait in \system.
Hence, the more the isolation wait time increases (as we add more threads), the more hidden the flushing latency is.}
Second, the \emph{partially-ordered \emph{durMarkers}} are 
dramatically faster than SPHT's totally-ordered durability wait (at peak throughput, overheads of 77\% in \system and 313\% in SPHT). 
\jpb{ -- despite SPHT optimizes such wait by employing a 
sophisticated group commit scheme. }
Overall, the savings attained by \system's durability optimizations 
prevail over the performance penalty of the isolation wait.
\jpb{We emphasize that this advantage is observed even though the workload does not run any RO transactions, which represent the type of workload for which \system is most optimized.} 

Interestingly, 
both variants of \system abort considerably more than SPHT.
In fact, in SPHT, a thread spends most of its time in the durability phase, which is performed out of the context of HTM;
whereas, in \system, most time is spent in the isolation wait, thus inside HTM.
Consequently, not only SPHT has fewer simultaneous active hardware transactions, but each has a shorter vulnerability window.
Both factors lead to higher conflict aborts in \system.
Still, 
the savings that \system attains in its durability phase (as previously discussed) 
also outweigh the penalty of higher abort rates.

 
When comparing the two DUMBO variants, it is perhaps surprising that \system-opa achieves a higher
throughput than \system-SI, despite having higher abort
rates. Analyzing abort  codes and time spent
on rolled-back transactions (available in our extended technical report \cite{anonymous-tech-report}) shows that \system-opa, by detecting conflicts earlier, spends considerably less time on rolled-back transactions. These savings outweigh the costs associated
of increases SGL acquisitions, suggesting that \system-opa may be more competitive in high-contention workloads with small transactions.

Regarding Pisces, it suffers from the overheads of the synchronization instrumentation on memory accesses 
(especially transactional writes).
Not surprisingly, it is the worst-performing persistent alternative.

The next workload, \emph{delivery}, 
is  extremely prone to capacity aborts on HTM (moderately high write footprints and very high read footprints).
In fact, with 1 thread, all solutions that run update transactions in full HTM (\system-Opa, SPHT, HTM) 
exhibit a severe capacity abort rate (81\%).
At higher thread counts, the scenario is worsened since contention induces transactional conflicts. 
%
In contrast, since \system-SI provides update transactions with unlimited reads, 
it exhibits substantially lower capacity abort rates 
(it still aborts nearly half of transactions due to write capacity).
Therefore, \system-SI attains dramatic throughput gains when compared to the HTM-based alternatives.
Such gains even 
conceal the costs of \system-SI's isolation wait.
%

Pisces is the only solution that enables unlimited reads as well as unlimited writes.
This allows Pisces to outperform the solutions based on full HTM (up to 12 threads).
However, the costly access instrumentation of an PSTM approach make it loose to \system-SI by a substantial margin.

For space limitations, we omit the \emph{neworder} workload from this individual analysis.
We consider it next, when we study mixed workloads.
An individual analysis of the \emph{neworder} workload would reveal very similar observations to the ones that 
we have drawn for the \emph{payment} workload above.

\subsection{Processing mixed workloads}
\label{sec:eval:mixed}

\begin{figure}[t]
	\centering
        $\quad\;$\includegraphics[width=0.4\textwidth]
        {ThroughputLegend_Hor.pdf} 

        \begin{tabular}{c c}
	\hspace{-2mm}
 \includegraphics[width=0.24\textwidth]
	{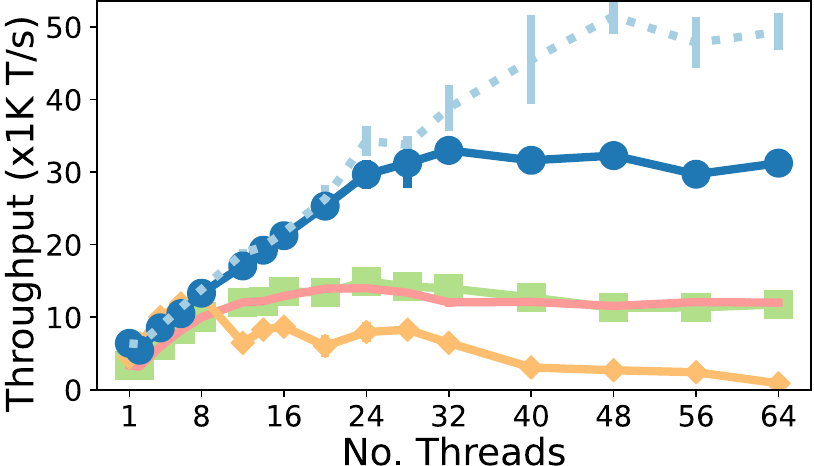}
        &
	\hspace{-3mm}
 \includegraphics[width=0.22\textwidth]
	{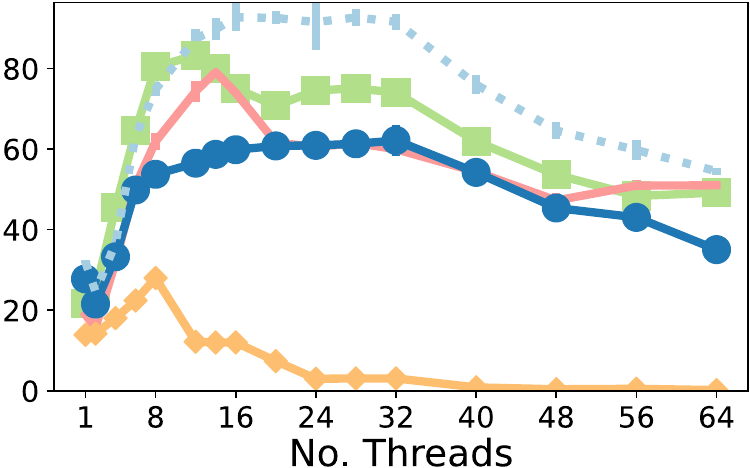}
	\\
        \end{tabular}

	\begin{tabular}{c c}
	\hspace{-2mm}
 \includegraphics[width=0.235\textwidth]
	{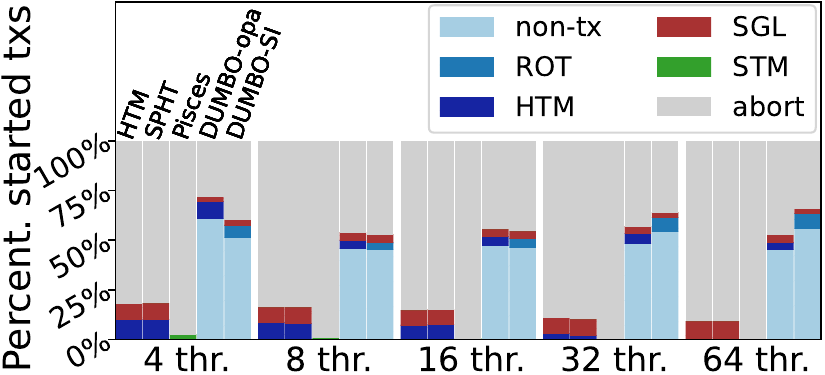}
	&
	\hspace{-3mm}
 \includegraphics[width=0.225\textwidth]
	{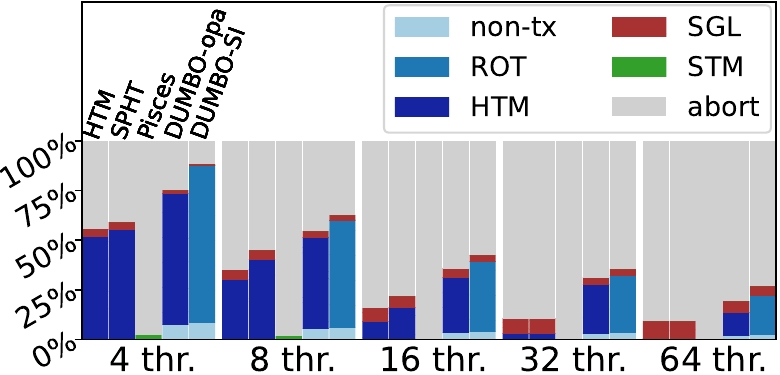}
	\\
    \end{tabular}
    \begin{tabular}{c c}
    \hspace{-2mm}
    \includegraphics[width=0.24\textwidth]
	{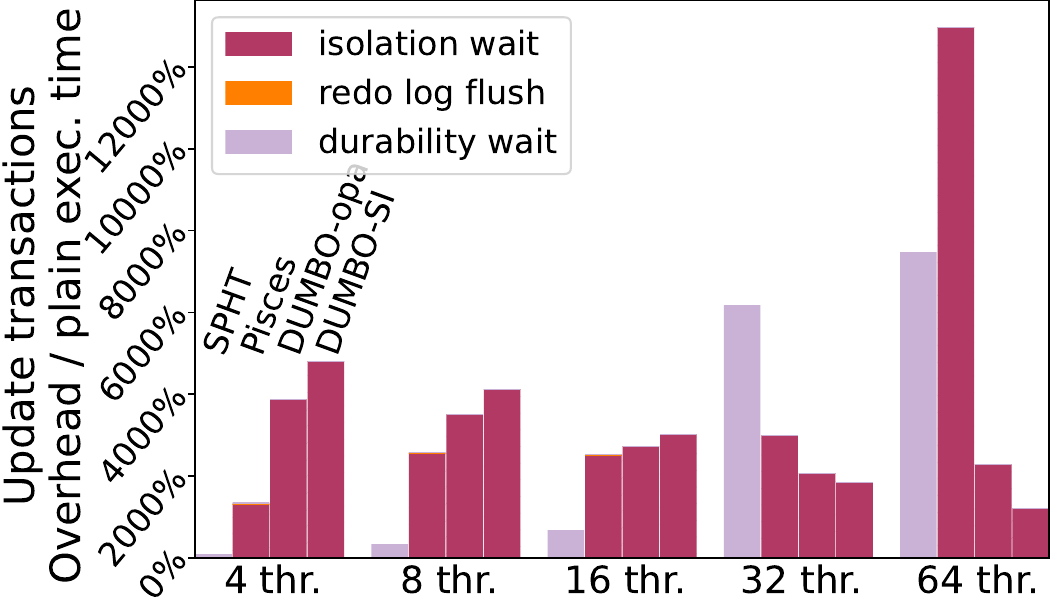}
	&
	\hspace{-2.9mm}
 \includegraphics[width=0.22\textwidth]
	{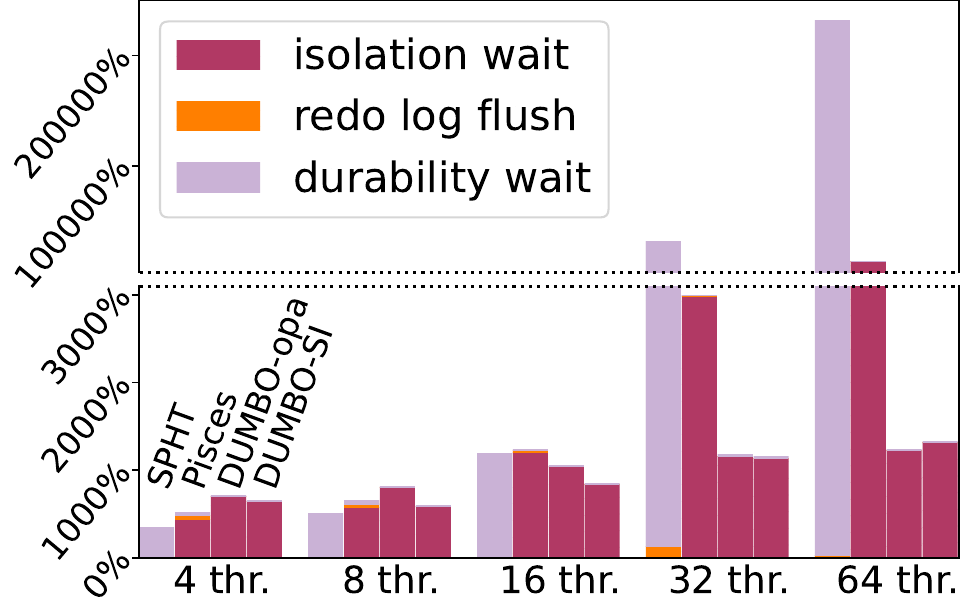}
	\\
    \end{tabular}

    \begin{tabular}{c c}
\hspace{-2mm}
\includegraphics[width=0.24\textwidth]
	{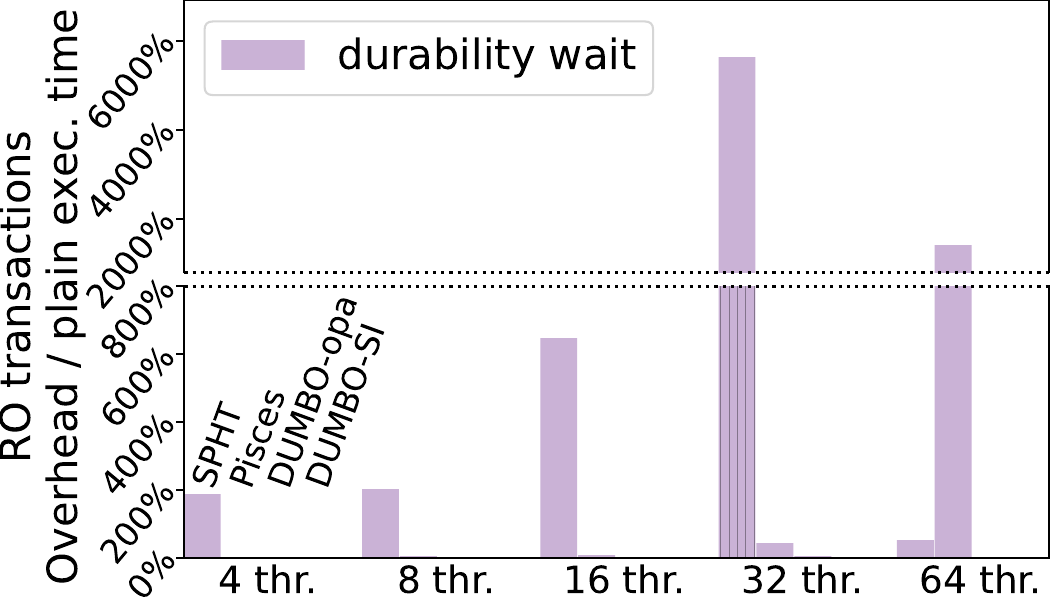}
	&
	\hspace{-7mm}
 \includegraphics[width=0.22\textwidth]
	{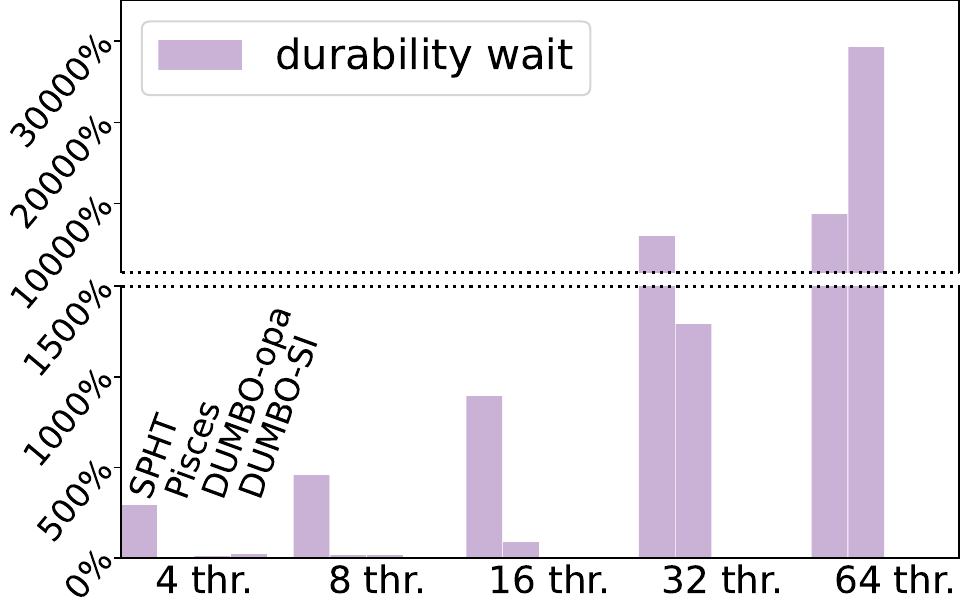}
	\\
    	\small{Read-dominated} & 
     \small{Standard update-dominated mix} \\
	\end{tabular}
 \vspace{-4mm}
	\caption{Throughput, transaction outcomes and overhead breakdown with mixed TPC-C workloads}
	\label{fig:tpcc_mixed}
\end{figure}

Figure \ref{fig:tpcc_mixed} presents our results with mixed workloads. 
We start by considering a read-dominated workload in Figure \ref{fig:tpcc_mixed}, in which 85\% of transactions are 
RO  (uniformly selected among \emph{stocklevel} or \emph{orderstatus}), and the remaining ones are from update types (\emph{delivery}, \emph{payment} 
or \emph{neworder}). 
%
For the first time, we have RO transactions that, in their durability wait, actually need to wait for update transactions. 
RO transactions in SPHT spend a substantial portion of time in the durability wait (up to $\approx 10\times$ the latency of the plain 
execution) until concurrent update transactions become durable (or abort). 
\jpb{which is in line with the results in \S\ref{sec:intro}, }
In contrast, the \emph{pruned RO durability wait} of \system is practically cost-less. 
A typical call to the corresponding routine finds the state array cached in L1
and does not find any transaction to wait for.



A collateral effect can be seen in the isolation wait latency of \system. 
Since the minority of update transactions now have to wait for a majority of RO transactions, the relative cost of the isolation wait 
grows so much that it overcomes the durability overheads of SPHT.
Even though the durability optimizations of \system  reduce the durability overheads to marginal levels,
the net balance is that update transactions in \system incur a higher overhead than in SPHT.
Still, from the  perspective of global throughput, this shortcoming is a means to achieve a greater good.

Pisces keeps up with \system's pace up to 8 threads and achieves throughput that is competitive with SPHT, 
before contention effects start manifesting more eminently on update transactions, hindering the overall scalability.

Finally, we look at the opposite extreme, an update-dominated mixed workload where 85\% transactions are either \emph{payment} or 
\emph{neworder}, and the remaining 15\% are from the other transaction types. 
 This is the standard mix mentioned in the TPC-C specifications \cite{tpcc}.
This workload is not favorable to \system, not just because RO transactions only represent 10\%, but also since the
transactions prone to capacity aborts only account to 15\%.
Still, \system-SI  manages to be the most competitive solution. 
Relatively to SPHT and Pisces, this advantage is explained by same factors that we have observed with \emph{payment} (\S\ref{sec:eval:upd}), 
which also affect \emph{neworder} -- i.e., the two dominant transactions.
Furthermore, \system's RO durability wait yields similar savings as in the read-dominated mix.
Comparing \system-SI with \system-opa, the less frequent capacity aborts in \system-SI (which grants unlimited reads to 
both kinds of transactions) is the decisive factor behind its throughput advantage over \system-opa. 

\begin{figure}[t] 
	\centering
	\includegraphics[width=0.30\textwidth]{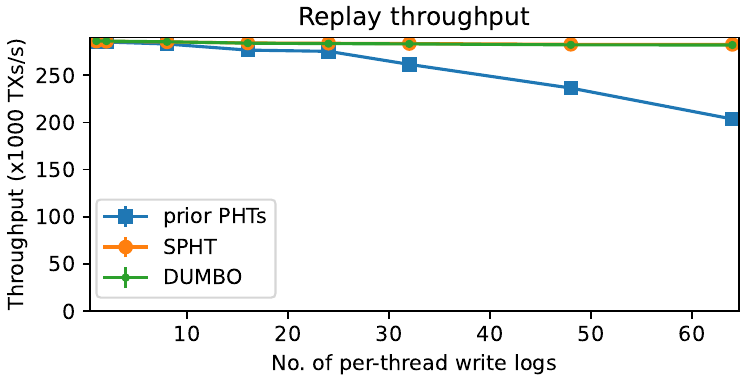}
  \vspace{-1mm}
	\caption{Log replay throughput of PHT alternatives}
	\label{fig:replay_cost}
\end{figure}	

\subsection{Log replay}
\label{sec:eval:logreplay}


We now compare the performance of the Log Replayer (LR)  of 
\system with SPHT  and
the LR of prior PHTs (cc-HTM~\cite{giles2017ISMM}, DudeTM~\cite{LiuASPLOS2017}, NV-HTM~\cite{CastroIPDPS2018}).
We use the same methodology as in SPHT's paper \cite{SPHT}.
First, we run a synthetic application with 100\% update transactions to prefill the per-thread logs.
Transactions issue between 1 and 20 writes (the number of writes and their accessed locations  are chosen uniformly at random), 
the log size and heap size is 128MB, and we vary the number of working threads.

Next, we halt the application and run the LR to fully replay the logs  and  measure the replay throughput. 
For space limitations, we only study the case where the LR runs in a single thread, and we disable filtering of duplicated writes \cite{CastroIPDPS2018}.
Using multi-threaded LR and duplicate  filtering techniques is compatible with \system's LR, 
and does not change the main conclusions that we draw next.

Figure~\ref{fig:replay_cost} presents the throughput of the different replay schemes. 
The LR of prior PHTs struggles with an increasingly larger amount of working threads. This is caused by the need to scan, after replaying one transaction, every per-thread log to find the next transaction to replay. 
In contrast, SPHT and \system prevent the scanning bottleneck, albeit by different means.
In SPHT, this is accomplished with its \emph{log linking} technique \cite{SPHT}; in \system, via our novel LR scheme based on the global \emph{durMarker} array.
As shown in Figure \ref{fig:replay_cost}, both LR schemes perform comparably and, most importantly, remain efficient even at high (worker) threads counts.

\section{Discussion}
\label{sec:discussion}

~\\\noindent\textbf{Limitations} 
Similarly to any prior work based on 
off-the-shelf HTM, \system does not circumvent the limited write capacity of HTM, 
so it is not suited to large write footprints.
In opacity mode (\system-opa), unlimited reads are only provided to those transactions that the program
explicitly flags as RO when invoking \texttt{beginTx}, which can be an issue with
legacy programs. In contrast, \system-SI transparently grants unlimited reads to any kind of transaction.

~\\\noindent\textbf{On the power of suspending access tracking in  HTM.}
 Supporting suspend/resume access tracking in HTM is well-known to incur higher complexity and cost in
 the design of the cache coherence protocol of a processor \cite{5695524}.
From a hardware-software co-design perspective, 
the hardware manufacturers will consider high-cost features only if they observe a 
clear demand by software developers. 
However, among academic research, we are only aware of a handful of works that 
exploit advanced suspend/resume features of HTM. 
To our knowledge, among software practitioners, the adoption of such advanced features is also
negligible.
We believe that our results contribute to a better awareness of the potential of suspend/resume 
among the research community and the microprocessor manufacturers. 
Our results also provide evidence that kernel-assisted
suspend/resume implementations, while reducing hardware complexity, can still enable important gains, 
even in workloads that make intense use of such instructions.

~\\\noindent\textbf{Adapting our techniques to Intel TSX.}
\system's design is incompatible with the latest generations of Intel TSX, which only enable
suspending load tracking, but not \emph{any-access} tracking.
More precisely, the optimizations that 
we propose for the durability phase of update transactions (see \S\ref{sec:durphase}) intrinsically 
depend on  any-access tracking suspension. 

However, the modules of \system that benefit RO transactions (with unlimited reads and 
the pruned durability wait) can be adapted to an HTM that 
only supports load tracking suspension, as Intel TSX. 
To accomplish this, we can adapt the isolation wait to only scan (and spin on) the state of RO transactions.
Consequently, update transactions no longer need to externalize their state transitions 
from within the suspend-resume window; therefore, only load tracking needs to be suspended in that window.
Update transactions still need to advertise that they are entering a non-durable state, but this
can be done just after access tracking is resumed (i.e., just before the HTM commit).
This solution would retain the two major benefits that \system provides to RO transactions.
Understanding the design and performance trade-offs of this approach in a real Intel TSX system is 
left for future work.

\section{Concluding remarks}
\label{sec:conclusions}

We proposed \system, a new design that
eliminates two  fundamental bottlenecks of durable RO transactions on HTM. 
At its core, \system exploits 
instructions that some contemporary HTMs
provide to suspend (and resume) transactional access tracking.
Our experimental results show that \system can outperform state of the art alternatives
in both PHT and PSTM, by up to 4.0$\times$. 

\section*{Acknowledgements}

This work was supported by FAPESP (2018/15519-5, 2019/10471-7), and by FCT, Fundação para a Ciência e a Tecnologia, 
with references UIDB/50021/2020. 

We acknowledge Oregon State University's Open Source Lab for having kindly provided us with
access to the IBM POWER9 machine.

\clearpage

\bibliographystyle{plain}
\bibliography{references}

\begin{thebibliography}{10}

\bibitem{tpcc}
{Transaction Processing Performance Council: TPC-C Benchmark Revision 5.11.0}.
\newblock
  \url{http://tpc.org/tpc\_documents\_current\_versions/pdf/tpc-c\_v5.11.0.pdf},
  February 2010.

\bibitem{intelManual}
{\em Intel 64 and IA-32 Architectures Software Developer's Manual - Volume 3},
  2023.

\bibitem{lazypersistency18}
Mohammad Alshboul, James Tuck, and Yan Solihin.
\newblock Lazy persistency: A high-performing and write-efficient software
  persistency technique.
\newblock In {\em 2018 ACM/IEEE 45th Annual International Symposium on Computer
  Architecture (ISCA)}, pages 439--451, 2018.

\bibitem{anonymous-tech-report}
anonymous.
\newblock Technical report to be disclosed once the paper is published.
\newblock Technical report, anonymous, 2025.

\bibitem{ARMHTM}
ARM.
\newblock Arm c language extensions (2021q2).
\newblock Online, july 2021.

\bibitem{avni2016}
Hillel Avni and Trevor Brown.
\newblock {Persistent hybrid transactional memory for databases}.
\newblock {\em Proceedings of the VLDB Endowment}, 10(4):409--420, nov 2016.

\bibitem{avni2015}
Hillel Avni, Eliezer Levy, and Mendelson Avi.
\newblock {Hardware Transactions in Nonvolatile Memory}.
\newblock In {\em LNCS 9363, 29th International Symposium on Distributed
  Computing}, volume 9363, pages 617--630. Springer, 2015.

\bibitem{BaldassinCSUR2021}
Alexandro Baldassin, Jo{\~a}o Barreto, Daniel Castro, and Paolo Romano.
\newblock {Persistent Memory: A Survey of Programming Support and
  Implementations}.
\newblock In {\em {ACM Computing Surveys Vol. 54, No. 7}}, pages 1--37, July
  2021.

\bibitem{Baldassin2020}
Alexandro Baldassin, Rafael Murari, João~P.L. de~Carvalho, Guido Araujo,
  Daniel Castro, João Barreto, and Paolo Romano.
\newblock Nv-phtm: An efficient phase-based transactional system for
  non-volatile memory.
\newblock volume 12247 LNCS, 2020.

\bibitem{Berenson1995SIGMOD}
Hal Berenson, Phil Bernstein, Jim Gray, Jim Melton, Elizabeth O’Neil, and
  Patrick O'Neil.
\newblock {A Critique of ANSI SQL Isolation Levels}.
\newblock In {\em {In Proceedings of the 1995 ACM SIGMOD International
  Conference on Management of Data (SIGMOD ’95}}, pages 1--10, June 1995.

\bibitem{SPHT}
Daniel Castro, Alexandro Baldassin, Jo{\~a}o Barreto, and Paolo Romano.
\newblock {SPHT}: Scalable persistent hardware transactions.
\newblock In {\em 19th USENIX Conference on File and Storage Technologies (FAST
  21)}, pages 155--169. USENIX Association, February 2021.

\bibitem{CastroIPDPS2018}
Daniel Castro, Paolo Romano, and Jo{\~a}o Barreto.
\newblock {Hardware Transactional Memory Meets Memory Persistency}.
\newblock In {\em {2018 IEEE International Parallel and Distributed Processing
  Symposium (IPDPS)}}, pages 368--377, 2018.

\bibitem{CASTRO201963}
Daniel Castro, Paolo Romano, and João Barreto.
\newblock Hardware transactional memory meets memory persistency.
\newblock {\em Journal of Parallel and Distributed Computing}, 130:63--79,
  2019.

\bibitem{castro2017MASCOTS}
Daniel Castro, Paolo Romano, Diego Didona, and Willy Zwaenepoel.
\newblock An analytical model of hardware transactional memory.
\newblock In {\em 2017 IEEE 25th International Symposium on Modeling, Analysis,
  and Simulation of Computer and Telecommunication Systems (MASCOTS)}, pages
  221--231, 2017.

\bibitem{Cerone2016PODC}
Andrea Cerone and Alexey Gotsman.
\newblock {Analysing Snapshot Isolation}.
\newblock In {\em {PODC '16: Proceedings of the 2016 ACM Symposium on
  Principles of Distributed Computing}}, pages 55--64, July 2016.

\bibitem{5695524}
Jaewoong Chung, Luke Yen, Stephan Diestelhorst, Martin Pohlack, Michael
  Hohmuth, David Christie, and Dan Grossman.
\newblock Asf: Amd64 extension for lock-free data structures and transactional
  memory.
\newblock In {\em 2010 43rd Annual IEEE/ACM International Symposium on
  Microarchitecture}, pages 39--50, 2010.

\bibitem{Coburn2011NVHeaps}
Joel Coburn, Adrian~M. Caulfield, Ameen Akel, Laura~M. Grupp, Rajesh~K. Gupta,
  Ranjit Jhala, and Steven Swanson.
\newblock Nv-heaps: Making persistent objects fast and safe with
  next-generation, non-volatile memories.
\newblock {\em SIGARCH Comput. Archit. News}, 39(1):105–118, mar 2011.

\bibitem{OptaneCXL}
Intel Corporation.
\newblock From intel® optane™ persistent memory to cxl.
\newblock
  \url{https://www.intel.com/content/www/us/en/products/docs/memory-storage/optane-persistent-memory-to-cxl-attached-memory.html},
  2022.
\newblock Accessed: 18 Oct, 2024.

\bibitem{Fekete2015}
Alan Fekete, Dimitrios Liarokapis, Elizabeth O'Neil, Patrick O'Neil, and Dennis
  Shasha.
\newblock {Making snapshot isolation serializable}.
\newblock In {\em {ACM Transactions on Database Systems, Vol. 30, No. 2}},
  pages 492--528, June 2015.

\bibitem{Felber2016EuroSys}
Pascal Felber, Paolo Romano, Alexander Matveev, and Shady Issa.
\newblock {Hardware Read-Write Lock Elision}.
\newblock In {\em {EuroSys '16: Proceedings of the Eleventh European Conference
  on Computer Systems}}, pages 1--15, April 2016.

\bibitem{FilipePPoPP2019}
Ricardo Filipe, Shady Issa, Jo{\~a}o Barreto, and Paolo Romano.
\newblock {Stretching the capacity of Hardware Transactional Memory in IBM
  POWER architectures}.
\newblock In {\em {PPoPP '19: Proceedings of the 24th Symposium on Principles
  and Practice of Parallel Programming}}, pages 107–--119, February 2019.

\bibitem{10.1145/3624062.3624175}
Yehonatan Fridman, Suprasad Mutalik~Desai, Navneet Singh, Thomas Willhalm, and
  Gal Oren.
\newblock Cxl memory as persistent memory for disaggregated hpc: A practical
  approach.
\newblock In {\em Proceedings of the SC '23 Workshops of The International
  Conference on High Performance Computing, Network, Storage, and Analysis},
  SC-W '23, page 983–994, New York, NY, USA, 2023. Association for Computing
  Machinery.

\bibitem{GencPLDI2020}
Kaan Gen\c{c}, Michael~D. Bond, and Guoqing~Harry Xu.
\newblock {Crafty: efficient, HTM-compatible persistent transactions}.
\newblock In {\em {PLDI 2020: Proceedings of the 41st ACM SIGPLAN Conference on
  Programming Language Design and Implementation}}, pages 59--74, June 2020.

\bibitem{giles2017ISMM}
Ellis Giles, Kshitij Doshi, and Peter Varman.
\newblock {Continuous Checkpointing of HTM Transactions in NVM}.
\newblock {\em Proceedings of the 2017 ACM SIGPLAN International Symposium on
  Memory Management -- ISMM 2017}, pages 70--81, 2017.

\bibitem{gileshtpm}
Ellis Giles, Kshitij Doshi, and Peter Varman.
\newblock Hardware transactional persistent memory.
\newblock In {\em Proceedings of the International Symposium on Memory
  Systems}, page 190–205, 2018.

\bibitem{Gu2019ATC}
Jinyu Gu, Qianqian Yu, Xiayang Wang, Zhaoguo Wang, Binyu Zang, Haibing Guan,
  and Haibo Chen.
\newblock Pisces: A scalable and efficient persistent transactional memory.
\newblock In {\em 2019 USENIX Annual Technical Conference (USENIX ATC 19)},
  pages 913--928, Renton, WA, July 2019. USENIX Association.

\bibitem{opacity}
Rachid Guerraoui and Michal Kapalka.
\newblock On the correctness of transactional memory.
\newblock In {\em Proceedings of the 13th ACM SIGPLAN Symposium on Principles
  and Practice of Parallel Programming}, PPoPP '08, page 175–184, 2008.

\bibitem{IssaDISC2017}
Shady Issa, Pascal Felber, Alexander Matveev, and Paolo Romano.
\newblock {Extending Hardware Transactional Memory Capacity via Rollback-Only
  Transactions and Suspend/Resume}.
\newblock In Andr{\'e}a~W. Richa, editor, {\em 31st International Symposium on
  Distributed Computing (DISC 2017)}, volume~91 of {\em Leibniz International
  Proceedings in Informatics (LIPIcs)}, pages 28:1--28:16, Dagstuhl, Germany,
  2017. Schloss Dagstuhl--Leibniz-Zentrum fuer Informatik.

\bibitem{Issa2018Middleware}
Shady Issa, Paolo Romano, and Tiago Lopes.
\newblock Speculative read write locks.
\newblock In {\em Proceedings of the 19th International Middleware Conference},
  page 214–226, 2018.

\bibitem{joshi18isca}
Arpit Joshi, Vijay Nagarajan, Marcelo Cintra, and Stratis Viglas.
\newblock {{DHTM}}: {{Durable Hardware Transactional Memory}}.
\newblock In {\em ISCA'18}, pages 452--465, June 2018.

\bibitem{atom}
Arpit Joshi, Vijay Nagarajan, Stratis Viglas, and Marcelo Cintra.
\newblock {ATOM}: Atomic durability in non-volatile memory through hardware
  logging.
\newblock In {\em 2017 IEEE International Symposium on High Performance
  Computer Architecture (HPCA)}, pages 361--372, 2017.

\bibitem{SAPHANA_HPCA14}
Tomas Karnagel, Roman Dementiev, Ravi Rajwar, Konrad Lai, Thomas Legler,
  Benjamin Schlegel, and Wolfgang Lehner.
\newblock Improving in-memory database index performance with intel®
  transactional synchronization extensions.
\newblock In {\em 2014 IEEE 20th International Symposium on High Performance
  Computer Architecture (HPCA)}, pages 476--487, 2014.

\bibitem{power9:suspendtrap}
The~Linux kernel~development community.
\newblock Transactional memory support.
\newblock
  \url{https://www.kernel.org/doc/html/v5.12/powerpc/transactional_memory.html},
  Last accessed 18 Oct, 2024.

\bibitem{KolliASPLOS2016}
Aasheesh Kolli, Steven Pelley, Ali Saidi, Peter~M Chen, and Thomas~F Wenisch.
\newblock {High-Performance Transactions for Persistent Memories}.
\newblock {\em Proceedings of the twenty first international conference on
  Architectural support for programming languages and operating systems --
  ASPLOS'16}, pages 399--411, 2016.

\bibitem{krish2020asplos}
R.~Madhava Krishnan, Jaeho Kim, Ajit Mathew, Xinwei Fu, Anthony Demeri,
  Changwoo Min, and Sudarsun Kannan.
\newblock Durable {{Transactional Memory Can Scale}} with {{TimeStone}}.
\newblock In {\em ASPLOS'20}, pages 335--349, March 2020.

\bibitem{le15ibm}
H.~Q. Le, G.~L. Guthrie, D.~E. Williams, M.~M. Michael, B.~G. Frey, W.~J.
  Starke, C.~May, R.~Odaira, and T.~Nakaike.
\newblock Transactional memory support in the {IBM} {POWER8} processor.
\newblock {\em {IBM} Journal of Research and Development}, 59(1):8:1--8:14,
  January 2015.

\bibitem{LiuASPLOS2017}
Mengxing Liu, Mingxing Zhang, Kang Chen, Xuehai Qian, Yongwei Wu, Weimin Zheng,
  and Jinglei Ren.
\newblock {DudeTM: Building Durable Transactions with Decoupling for Persistent
  Memory}.
\newblock In {\em {ASPLOS '17: Proceedings of the Twenty-Second International
  Conference on Architectural Support for Programming Languages and Operating
  Systems}}, pages 329--–343, April 2017.

\bibitem{10.1145/3582016.3582063}
Hasan~Al Maruf, Hao Wang, Abhishek Dhanotia, Johannes Weiner, Niket Agarwal,
  Pallab Bhattacharya, Chris Petersen, Mosharaf Chowdhury, Shobhit Kanaujia,
  and Prakash Chauhan.
\newblock Tpp: Transparent page placement for cxl-enabled tiered-memory.
\newblock In {\em Proceedings of the 28th ACM International Conference on
  Architectural Support for Programming Languages and Operating Systems, Volume
  3}, ASPLOS 2023, page 742–755, New York, NY, USA, 2023. Association for
  Computing Machinery.

\bibitem{samsung:cxl}
Samsung.
\newblock Samsung cxl solutions – cmm-h.
\newblock
  \url{https://semiconductor.samsung.com/news-events/tech-blog/samsung-cxl-solutions-cmm-h/},
  Last accessed 18 Oct, 2024.

\bibitem{durableopacity20}
Eleni Vafeiadi~Bila, Simon Doherty, Brijesh Dongol, John Derrick, Gerhard
  Schellhorn, and Heike Wehrheim.
\newblock {\em Defining and Verifying Durable Opacity: Correctness for
  Persistent Software Transactional Memory}, pages 39--58.
\newblock Springer, 06 2020.

\bibitem{VolosASPLOS2011}
Haris Volos, Andres~Jaan Tack, and Michael~M. Swift.
\newblock {Mnemosyne: Lightweight Persistent Memory}.
\newblock In {\em {ASPLOS XVI: Proceedings of the sixteenth international
  conference on Architectural support for programming languages and operating
  systems}}, pages 91--104, March 2011.

\bibitem{wang15cal}
Zhaoguo Wang, Han Yi, Ran Liu, Mingkai Dong, and Haibo Chen.
\newblock Persistent {{Transactional Memory}}.
\newblock {\em IEEE Comput. Archit. Lett.}, 14(1):58--61, January 2015.

\bibitem{morlog}
Xueliang Wei, Dan Feng, Wei Tong, Jingning Liu, and Liuqing Ye.
\newblock {MorLog}: Morphable hardware logging for atomic persistence in
  non-volatile main memory.
\newblock In {\em 2020 ACM/IEEE 47th Annual International Symposium on Computer
  Architecture (ISCA)}, pages 610--623, 2020.

\bibitem{KJI+21}
Kai Wu, Jie Ren, Ivy Peng, and Dong Li.
\newblock {ArchTM}: {Architecture-Aware}, high performance transaction for
  persistent memory.
\newblock In {\em 19th USENIX Conference on File and Storage Technologies (FAST
  21)}, pages 141--153. USENIX Association, February 2021.

\bibitem{10.1145/3492321.3519556}
Lingfeng Xiang, Xingsheng Zhao, Jia Rao, Song Jiang, and Hong Jiang.
\newblock Characterizing the performance of intel optane persistent memory: A
  close look at its on-dimm buffering.
\newblock In {\em Proceedings of the Seventeenth European Conference on
  Computer Systems}, EuroSys '22, page 488–505, New York, NY, USA, 2022.
  Association for Computing Machinery.

\bibitem{XIS21}
Yi~Xu, Joseph Izraelevitz, and Steven Swanson.
\newblock Clobber-nvm: Log less, re-execute more.
\newblock In {\em Proceedings of the 26th ACM International Conference on
  Architectural Support for Programming Languages and Operating Systems},
  ASPLOS 2021, pages 346–--359, New York, NY, USA, 2021. Association for
  Computing Machinery.

\bibitem{SpecPMT_ASPLOS23}
Chencheng Ye, Yuanchao Xu, Xipeng Shen, Yan Sha, Xiaofei Liao, Hai Jin, and Yan
  Solihin.
\newblock Specpmt: Speculative logging for resolving crash consistency overhead
  of persistent memory.
\newblock In {\em Proceedings of the 28th ACM International Conference on
  Architectural Support for Programming Languages and Operating Systems, Volume
  2}, ASPLOS 2023, page 762–777, New York, NY, USA, 2023. Association for
  Computing Machinery.

\bibitem{yoo13sc}
R.~M. Yoo, C.~J. Hughes, K.~Lai, and R.~Rajwar.
\newblock Performance evaluation of {I}ntel transactional synchronization
  extensions for high-performance computing.
\newblock In {\em Proceedings of the International Conference for High
  Performance Computing, Networking, Storage and Analysis}, pages 1--11,
  November 2013.

\end{thebibliography}

\end{document}